\documentclass[a4paper]{article}
\usepackage{graphicx}
\usepackage{./styles/myxspace}
\usepackage{./styles/mysidecaption}
\sidecaptionraggedtrue

\begin{document}

\vspace{1.0cm}
\begin{center}

{\bf {\huge High-Energy Neutrino Astronomy}  

\bigskip

{\Large A Glimpse of the Promised Land}
\footnote{talk given at the session of the Russian Academy of Science 
dedicated to Bruno Pontecorvo, Dubna, Sept.\,2013}
 
\vspace{0.5cm}

{\bf Christian Spiering}

\medskip

{\it DESY, Platanenallee, D-15738 Zeuthen} 

}
\end{center}

\vspace{1.0cm}

{\small 
\noindent
{\bf Abstract:}
In 2012, physicists and astronomers celebrated the hundredth anniversary
of the detection of cosmic rays by Viktor Hess. One year later, in 2013, there was
first evidence for extraterrestrial high-energy neutrinos, i.e. for a signal
which may contain key information on the origin of cosmic rays.
That evidence is provided by data taken with
the IceCube neutrino telescope at the South Pole. 
First concepts to build a detector of this kind have been discussed
at the 1973 International Cosmic Ray Conference. Nobody would
have guessed at that time that the march towards first discoveries
would take forty years, the biblical time of the march from Egypt to
Palestine. But now, after all, the year 2013 has provided us a first glimpse
to the promised land of the neutrino universe at highest energies.
This article sketches the
evolution towards detectors with a realistic discovery potential,
describes the recent relevant results obtained with the IceCube 
and ANTARES neutrino telescopes and tries a look into the future. 
}

\section{Introduction}

The year 2012 marks the hundredth anniversary of the detection of
cosmic rays by Viktor Hess. As we know today,
cosmic rays consist essentially of protons and nuclei of heavier elements;
electrons contribute only at the percent level.  Since cosmic rays are
electrically charged, they are deflected by cosmic magnetic fields
on their way to Earth. Precise pointing -- i.e. astronomy -- is
only possible with electrically neutral, stable particles: electromagnetic
waves (i.e. gamma rays at the energies under consideration) and
neutrinos. High energy neutrinos,
with energies much beyond a GeV, must be emitted as a by-product
of collisions of charged cosmic rays with matter. Actually, only
neutrinos provide incontrovertible evidence for 
acceleration of hadrons since gamma rays may also stem from
inverse Compton scattering of accelerated electrons and other electromagnetic
processes. 
Moreover, neutrinos can escape much denser celestial environments
than light, therefore they can be tracers of processes which stay
hidden to traditional and gamma ray astronomy. At the same time, however, their
extremely low reaction cross section  makes their detection a challenge
and requires huge detectors.

This article follows the evolution towards IceCube, the first neutrino telescope with a 
realistic discovery potential, sketches the recent relevant results obtained with large 
neutrino telescopes and tries a look into the future. Recent reviews of the field
can be found in \cite{reviews,KS}, a detailed review of its history in \cite{history}.

\section{The idea}
\label{sec-idea}

The initial idea of neutrino astronomy beyond the solar system  
rested on two arguments: The first was the
expectation that a supernova stellar collapse in our galaxy would be accompanied
by an enormous burst of neutrinos in the 5-50 MeV range. The second was the expectation 
that fast rotating pulsars must accelerate charged particles in their Tera-Gauss magnetic
fields. Either in the source or on their way to Earth they must hit matter or
radiation fields and generate
pions with neutrinos as decay products of the pions.  

Today we write, for interactions with a photon gas:
\begin{equation}
\hspace{-10mm}
p + \gamma  \rightarrow \Delta^+ \rightarrow \pi^+ + \mbox{n}  \hspace{10mm} \mbox{and} \hspace{5mm}
\pi^+ \rightarrow \mu^+ + \nu_{\mu} \rightarrow
e^+ + \nu_e + \bar{\nu}_{\mu} +  \nu_{\mu}
\end{equation}

The resulting neutrino flavor ratio is approximately $\nu_e:\nu_\mu:\nu_\tau= 1:2:0$ 
at the sources. Neutrino oscillation turns this into a ratio of 
$1:1:1$ upon arrival at Earth.

Since this talk was given at the occasion of Bruno Pontecorvo's 100th birthday, it is worth to remind
that equation (1) reflects his concept of generating a second (muonic) neutrino
via pion decay \cite{Pontecorvo-1}, whereas the transformation from a flavor ratio 1:2:0 to 1:1:1 
goes back to his idea of neutrino oscillations \cite{Pontecorvo-2,Pontecorvo-3}. Indeed,
much of what we do in high-energy neutrino astronomy
is deeply rooted in some of the basic ideas of this visionary physicist.

The first thoughts to detect cosmic high energy neutrinos underground or underwater
date back to the late fifties.  In 1960, 
Kenneth Greisen and Frederick Reines discussed the motivations and prospects for such detectors
\cite{Greisen-1960,Reines-1960}.
In the same year, on the 1960 Rochester Conference, Moisei Markov published his
groundbreaking idea   
{\it "...to install detectors deep in a lake or a sea and  to determine the 
direction of  charged particles with the help of Cherenkov radiation"}  \cite{Markov-1960}.
This appeared to be the only way to reach detector volumes beyond the scale of
$10^4$ tons.

\section{DUMAND: start of the adventure}
\label{s-Dumand}

\begin{figure}[ht]
\begin{center}
\includegraphics[width=12cm] {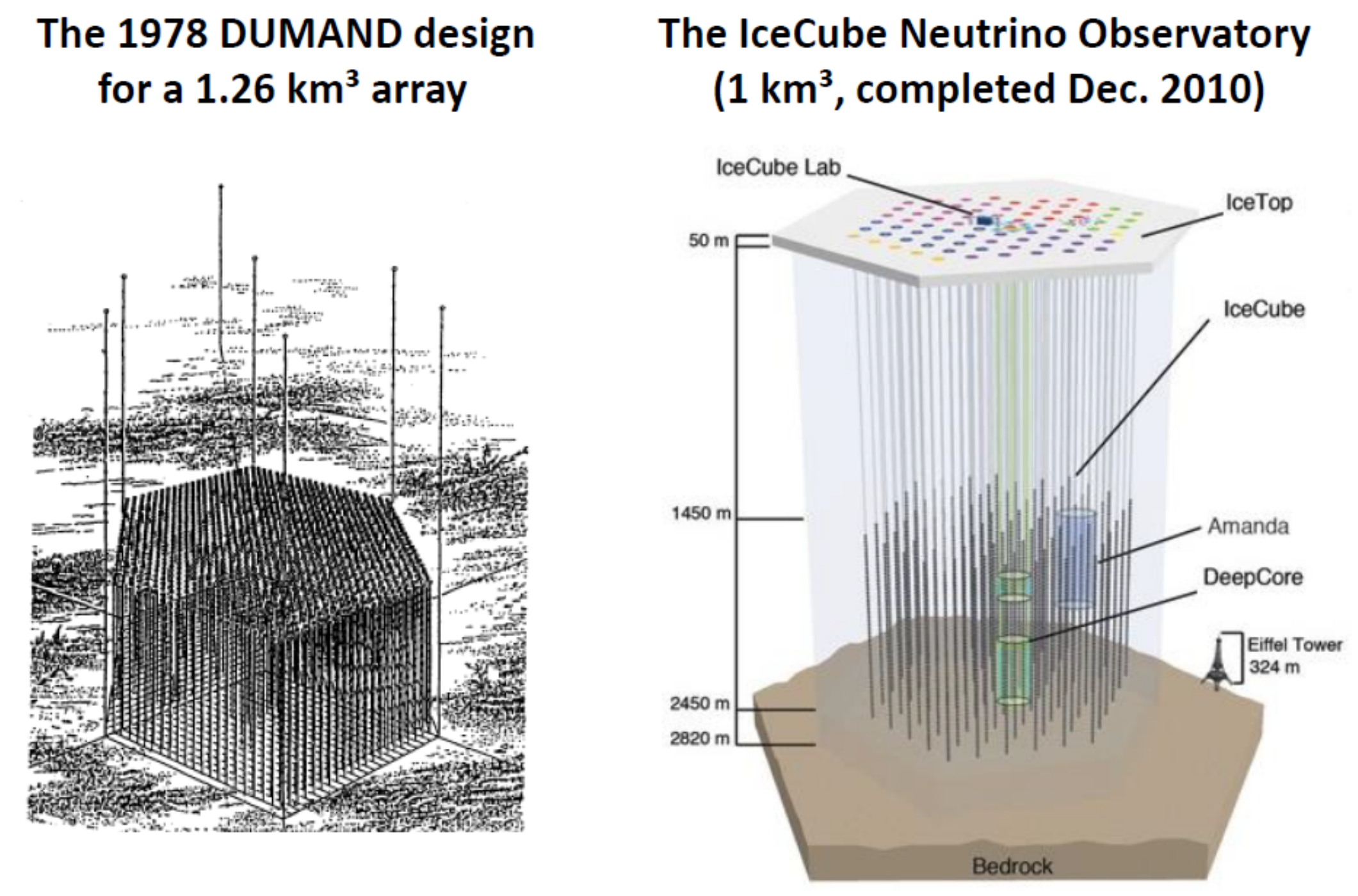}
\end{center}
\caption{  
A cubic kilometer detector: From dream to reality. 
The  DUMAND configuration conceived in 1978 (left) and
the IceCube detector completed in 2010 (right).
}  
\label{DUMAND}
\end{figure}

The real march towards underwater neutrino telescopes started forty years ago
at the 1973 International Cosmic Ray Conference.
There, a small group of physicists from the USA, Japan and Russia 
discussed for the first time building such a device:
the Deep Underwater Muon and Neutrino Detector (DUMAND).
The original design from 1978 envisaged an array of about 20\,000 photomultipliers (PMs)
spread over a 1.26 cubic kilometer volume (Fig.\ref{DUMAND}, left). 
The PMs would record arrival time and
amplitude of Cherenkov light emitted by muons or particle cascades.  
The size of the array was based {\it "... on relatively scant information
on the expected neutrino intensities and was difficult to justify in detail; the general
idea was that neutrino cross section are small and high-energy neutrinos are
scarce, so the detector had better be large."} \cite{DUMAND-Roberts}.

During the 1960s, no predictions or serious estimates for neutrino fluxes from 
cosmic accelerators had been published. Actually, many of the objects nowadays considered
as top candidates for neutrino emission (quasars, pulsars, X-ray binaries, gamma ray bursts) were discovered only in the sixties and seventies.   The situation changed in the 1970s,
when these objects were identified as possible neutrino emitters,
although still with highly uncertain flux predictions.  

First DUMAND discussions 
at the 1978 DUMAND workshop \cite{DUMAND-1978} 
focused on neutron star binary systems as point
sources of high energy neutrinos (specifically
Cygnus X-3), 
to neutrino signals from apparent
sources of TeV-$\gamma$-ray (none of them significant at that time!)
and to diffuse fluxes. 

A large number of papers on expected neutrino fluxes was published during the 1980s. The   
fluxes were found to depend strongly {\it a)} on the energy spectrum of the $\gamma$-ray
sources which could only be guessed since the first uncontroversial TeV-$\gamma$ observation
was the Crab nebula in 1989 \cite{Whipple-1989}, and {\it b)} on the
supposed $\nu/\gamma$ ratio which depends on the unknown thickness of matter
surrounding the source. 

Confronted with the oceanographic and financial reality, the size
of DUMAND was reduced in several steps down to a
216-PM version (DUMAND-II), to be deployed close to 
Hawaii at a depth of 4.8\,km.  It took more than three decades from
the 1978 DUMAND design to the actual completion of a cubic kilometer detector: the
IceCube Neutrino Observatory at the South Pole (see Fig.\ref{DUMAND}, right)!

The pessimistic and optimistic numbers for signal events given in 
the 1988 DUMAND proposal \cite{DUMAND-Project} differed by 2-4 orders of
magnitude and left it open whether DUMAND-II would be able to detect
neutrino sources or whether this would remain the realm of a future cubic kilometer
array.   In 1990, Venjamin Berezinsky estimated the necessary array size
to detect extragalactic sources as 0.1-1.0 km$^2$ size \cite{Berezinsky-1990}. 
DUMAND-II, with 25\,000 m$^2$ area, fell just below these values.

In 1987, the DUMAND collaboration operated a test string for some hours from a 
vessel and measured muon intensity as a function of depth 
\cite{DUMAND-Babson}.  One year later the DUMAND-II proposal
was submitted and another six years later a first full-scale string
deployed. Due to leakage problems the communication 
to shore failed. The recovered string was analyzed 
and a redeployment prepared.
But in spite of the remarkable progress compared to ocean technology at that time,
the risk aversion of funding agencies 
led to a termination of DUMAND in 1995. 

From now on
the goal to begin high energy neutrino astronomy was 
carried forward at the South Pole, in the Mediterranean Sea and 
in Lake Baikal.

\section{The long road: from NT200 to IceCube}
\label{s-longroad}

\subsection{NT200 in Lake Baikal}
\label{s-NT200}

In 1980, Alexander Chudakov proposed to use the deep water of Lake Baikal in Siberia
as the site for a "Russian DUMAND".  In late Winter the lake
is covered by a thick ice layer which allows  deploying  
underwater equipment without any use of ships. 
First shallow-site experiments with small PMTs started in 1981,
and soon a site in the Southern part of Lake Baikal,
3.6\,km to shore and at a depth of  about 1370\,m was identified as the
optimal location for a detector.
In 1984 and 1986, first stationary strings were deployed and
recorded downward moving muons \cite{Baikal-1984}. 

In 1989, a preliminary version of what later was called the NT200 project 
was
developed, an array comprising 192 optical modules at
8 strings \cite{Baikal-Project}. The volume of NT200 was only twice
that of Super-Kamiokande (or about $10^{-4}$\,km$^3$), but the possibility to see bright signals 
emerging outside the geometrical volume made the detector much more sensitive for
some high-energy processes.

\begin{figure}[ht]
\sidecaption
\includegraphics[width=4.5cm]{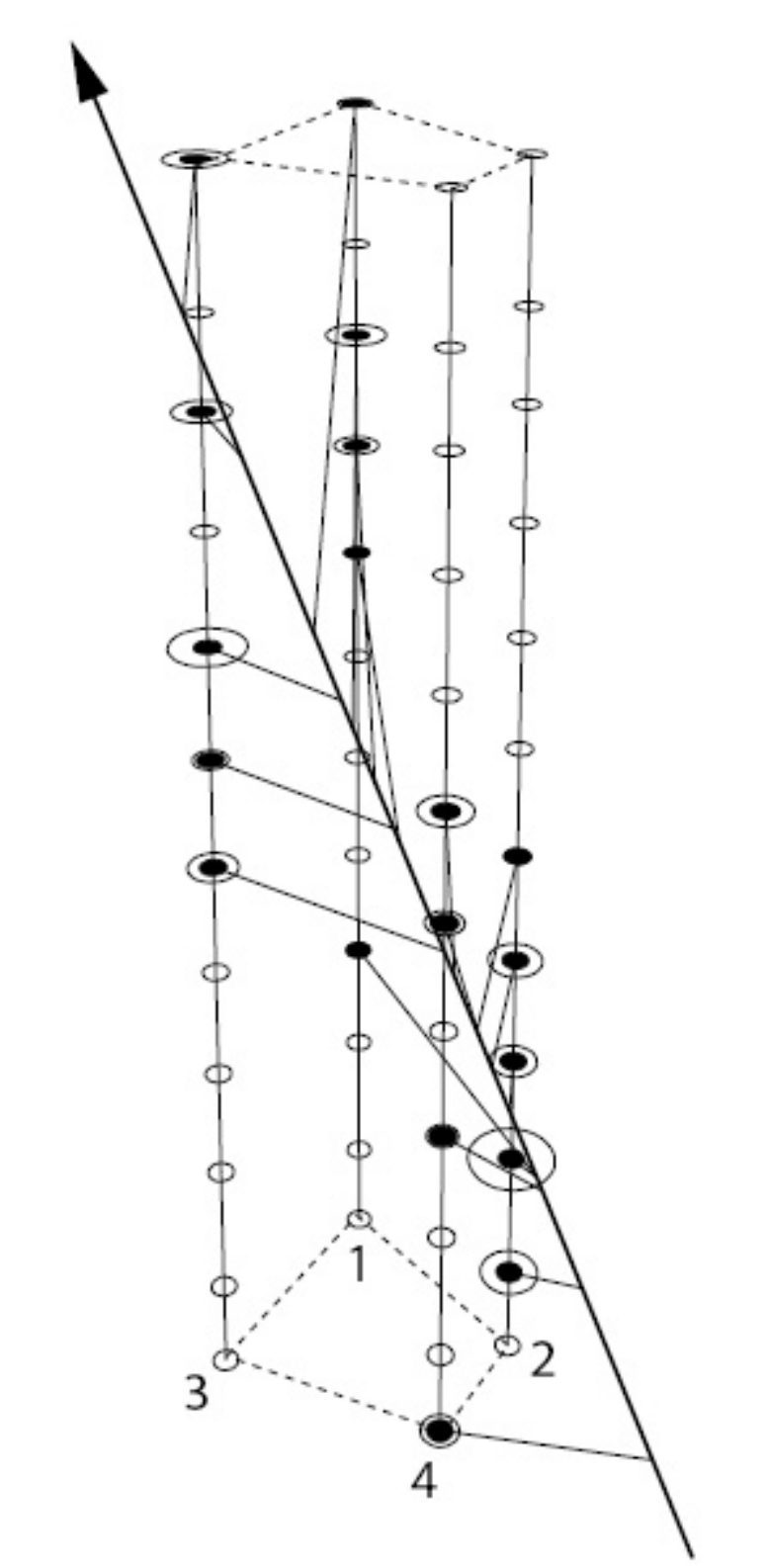}
\caption{   
One of the first upward moving
muons from a neutrino interaction recorded with the 4-string stage of the
Baikal detector in 1996 \cite{Baikal-atm}. The Cherenkov light from the
muon is recorded by 19 channels.} 
\label{Baikal}
\end{figure}

NT200 started with the deployment of a 3-string array \cite{NT-36} with 36 optical modules
in 1993.
The first two upward moving muons, i.e. neutrino candidates, were 
separated from the 1994 data.
In 1996, a 96-OM array with four NT200 strings was 
operated \cite{Baikal-atm} and provided the the first  textbook neutrino events 
like the one shown in  Fig.\ref{Baikal}.

NT200 was completed in April 1998 and has been taking data since then.

\subsection{AMANDA}
\label{ss-Amanda}
In 1988, Francis Halzen and John Learned 
released a paper "High energy neutrino detection in deep
Polar ice" \cite{Halzen-1988}.  This spectacular idea marked the
starting point for AMANDA (Antarctic Muon And Neutrino Detection Array).
Holes for the PMs were proposed to be drilled into the ice with
pressurized hot water.
After tests in Greenland and at the South Pole, in 1993/94 a first array
with 80 PMs on four strings was deployed at the South Pole
-- however at a too shallow depth, where the ice is still very bubbly
\cite{Amanda-shallow-1995}.  
In 1995/96, a  second 4-string array was installed at 
1500--2000\,m depth where the bubbles have disappeared.
It took some time to proof that the ice quality was indeed sufficient for
reconstruction, but in 1996 also AMANDA could provide the first
clear upward going muon events from neutrino interactions
\cite{Amanda-1999}.
The array was upgraded
stepwise until January 2000 and eventually comprised 19 strings with a total of
677 PMs.

\begin{figure}[ht]
\sidecaption
\includegraphics[width=8cm]{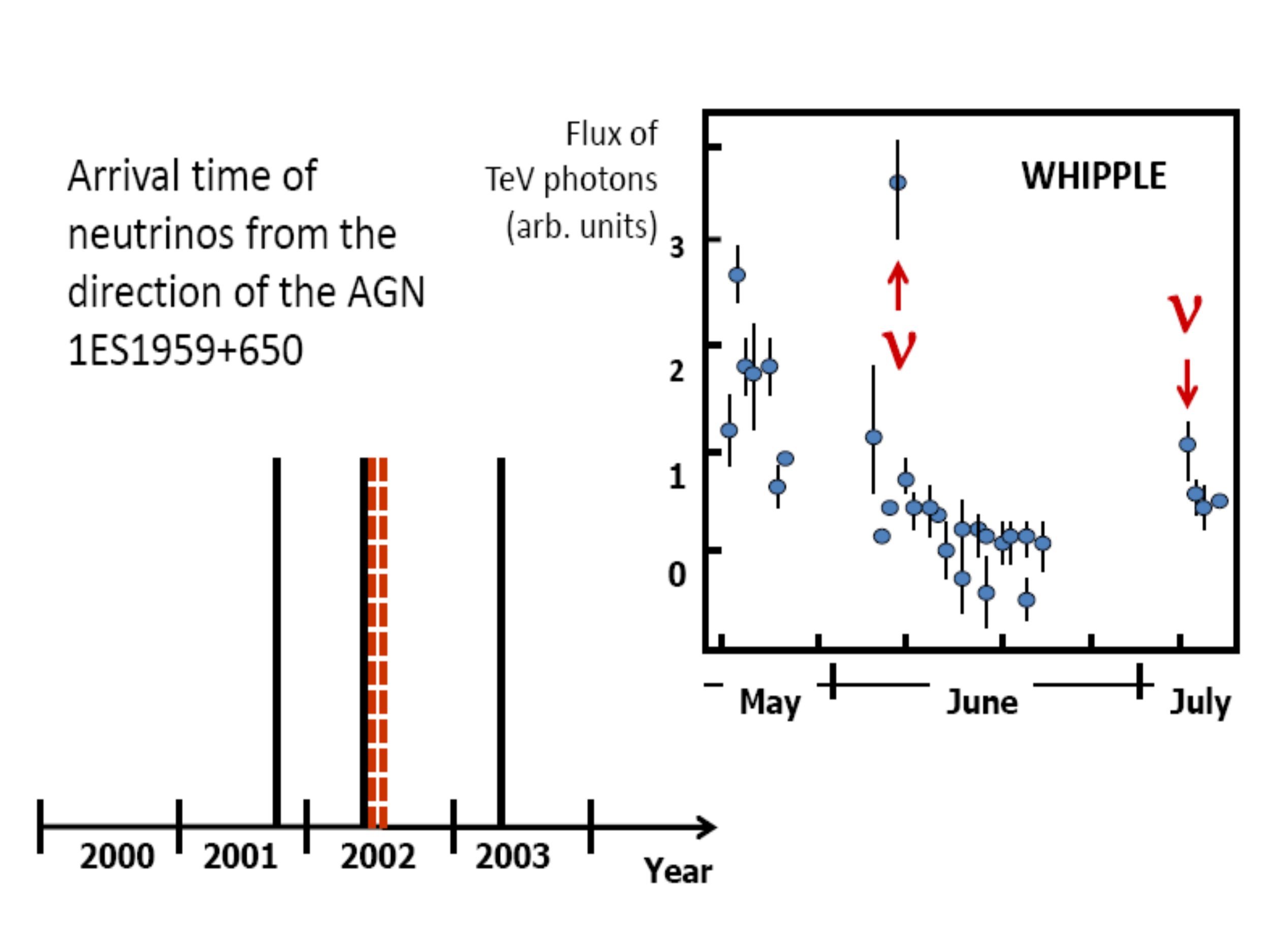}
\caption{   
The "curious" coincidence of neutrino events from the direction of
an AGN with gamma flares from the same source. The second and third
of the three events recorded in 2002 (dashed) coincide within about one
day with peaks seen by the Whipple telescope. 
\label{ES}
}
\end{figure}

AMANDA was switched off in April 2009, after 9 years of data
taking in its full configuration.  It  provided 6595 
atmospheric neutrinos, several important upper limits, but
no clear indication of any extraterrestrial neutrino signal. 
Actually, there was one short moment of hope.  While analyzing
in 2005 the data taken from 2000-2003, five events where identified from the direction of
the Active Galaxy 1ES1959+650. Interestingly, three of them came within 66~days 
in 2002 \cite{Markus-Thesis}. Two of
these three neutrinos were coinciding within about a day with gamma-ray flares
observed by the gamma-ray telescopes HEGRA and Whipple
 -- see Fig.\,\ref{ES}. Moreover, one of these two flares was not
accompanied by an X-ray flare, a so-called "orphan flare", which one would
expect for a hadron flare where the X-ray flux accompanying electron flares is
absent. This result was quickly followed by two theoretical papers, one claiming
that the corresponding neutrino flux would not fit any reasonable assumption on
the energetics of the source \cite{Reimer-2005}, the other claiming that
scenarios yielding such fluxes were conceivable \cite{Halzen-Hooper-2005}. 
This curious gamma-neutrino coincidence initiated considerations to send alerts to
gamma-ray telescopes in case time-clustered neutrino events from a certain direction would
appear. Such alert programs are currently operating
between IceCube and the gamma-ray telescopes MAGIC (La Palma) and
VERITAS (Arizona) \cite{NToO}.

\subsection{Mediterranean Projects: NESTOR, ANTARES, NEMO}
\label{s-MedTerr}

The exploration of the Mediterranean Sea as a site for an underwater
neutrino telescope started in 1989 when
Russian physicists measured, from a ship, the muon counting
rate as a function of depth \cite{Denyenko-91}.
In July 1991, a Greek/Russian collaboration 
deployed a hexagonal structure with 14 PMs
down to a depth of 4100\,m, at a site close to
Pylos at the coast of the Peloponnesus. 
This was the start of the NESTOR project \cite{nestor-1994}. 
NESTOR was conceived to consist of  seven towers, each comprising
12 hexagonal floors, spread over a volume of 0.1\,km$^3$. 
In 2004, a single prototype floor was deployed and operated for
about one month \cite{nestor-test}. Its operation was terminated 
due to a failure of the cable to shore.
Data taken with this prototype demonstrated the detector functionality 
and provided a measurement of the atmospheric muon flux \cite{nestor-atm-muons}.

The second Mediterranean project is ANTARES. 
It started in the mid 1990s; a full proposal  
was presented in 1999 \cite{antares-proposal}. 
ANTARES consists of 12 strings, each carrying 25 PM-triplets.
With a geometrical volume of 0.01 km$^3$ it has almost the
same size as AMANDA. 
ANTARES was constructed between 2002 and 2008. It
has convincingly demonstrated that a detector with precise angular resolution
can be reliably operated in the deep sea \cite{antares-detector}. 

The latest of the Mediterranean attempts is NEMO \cite{nemo}. 
The project was launched in 1998, with the objective to study the
feasibility of a cubic kilometer detector, to develop corresponding
technologies and to identify a suitable site.
The location is at a depth of
3.5\,km, about 100\,km off the South-Eastern coast of Sicily.
Several prototypes of towers ("rope-ladders" of
tilted bars with PMs at each end) have been deployed and recorded downward moving
muons \cite{nemo-atmuons}. At present, 
eight towers are being build and planned to be deployed until
early 2016. Later, they will be integrated
in a large future Mediterranean detector, KM3NeT (see sect.\,\ref{s-future}).

\subsection{IceCube}
\label{s-Icecube}

IceCube \cite{Henrike,Francis-Spencer} 
consists of 5160 digital optical modules (DOMs) installed on 86 strings at
depths of 1450 to 2450\,m.
A string carries 60~DOMs with 10-inch PMs housed in glass spheres. 
Signals  are digitized in the DOM and sent to the surface via copper cables.
320~further DOMs are installed in IceTop, an array of detector stations on the
ice surface directly above the strings (see Fig.\,1). AMANDA,
initially running as a low-energy sub-detector of IceCube, was decommissioned in
2009 and replaced by DeepCore, a high-density sub-array of eight strings at large
depths (in the clearest ice layer) at the center of IceCube. 
DeepCore collects photons with about six times the efficiency of full
IceCube, due to its smaller spacing, the better ice quality and the
higher quantum efficiency of new PMs.
Together with the veto provided by IceCube, this results in a
threshold of about 10\,GeV and opened a new venue for oscillation physics
and indirect dark matter search.

A first, single IceCube string was
deployed in January 2005, six years after submission of
the initial proposal to NSF \cite{IceCube-Proposal-1999}.
The following seasons resulted in 8, 13, 18, 19, 20 and 7 strings, respectively. 
The last of 86 strings was deployed at Dec.\,18, 2010.

\bigskip

Finally, the idea of  a cubic-kilometer detector was realized!

\section{Results}
\label{s-results}

Large neutrino telescopes underwater and in ice would never have
been built without the primary goal of identifying and understanding 
cosmic accelerators. But actually they are multi-purpose devices,
with a shopping list of impressive length. They are used
to search for signatures of dark matter particles and other
exotic forms of matter,  like magnetic monopoles. They allow studying
neutrino oscillations and other questions of particle physics -- like
cross sections for neutrino interactions and heavy particle production at highest energies.
IceCube, in addition, has sensitivity to a phenomenon much below the nominal
energy threshold of these detectors: to MeV-neutrinos from a supernova
collapse. Such neutrinos are emitted in a $\simeq$10 second burst and lead to
slightly enhanced counting rates of all PMs.  Last but not least
charged cosmic rays can be studied -- either with the help of downward
going punch-through muons from air showers or, like in the case of
IceCube, by an air shower array installed at the surface (IecTop).

In the following, I will focus on the search for high-energy extraterrestrial neutrinos 
and on the study of neutrino oscillations with atmospheric neutrinos. It are
these frontiers where the most remarkable progress of the last 2 years has happened.
Recent results come from IceCube \cite{ICRC-IC} and ANTARES 
\cite{ICRC-ANT} while Baikal NT200 has provided important limits in the past, 
notably on diffuse fluxes of extraterrestrial neutrinos and on the flux of magnetic monopoles
\cite{Baikal-diff,Baikal-mon}. 


\subsection{Atmospheric neutrinos}    
\label{ss-atm}

Atmospheric neutrinos and muons are produced in cosmic-ray interactions in the
atmosphere. Up to energies of about 100\,TeV, their flux is dominated by 
neutrinos from pion
and kaon decays. The corresponding neutrinos are referred to as "conventional''
atmospheric neutrinos. Their spectrum follows approximately an $E^{-3.7}$ shape. 
At higher energies,
"prompt'' atmospheric neutrinos from the decay of charm and bottom particles
take over. These particles decay before having a chance for further
interactions. Therefore the resulting neutrinos closely follow the primary
cosmic ray power law spectrum with its $E^{-2.7}$ shape.

\begin{figure}[h]
\center{
\includegraphics[width=12cm]{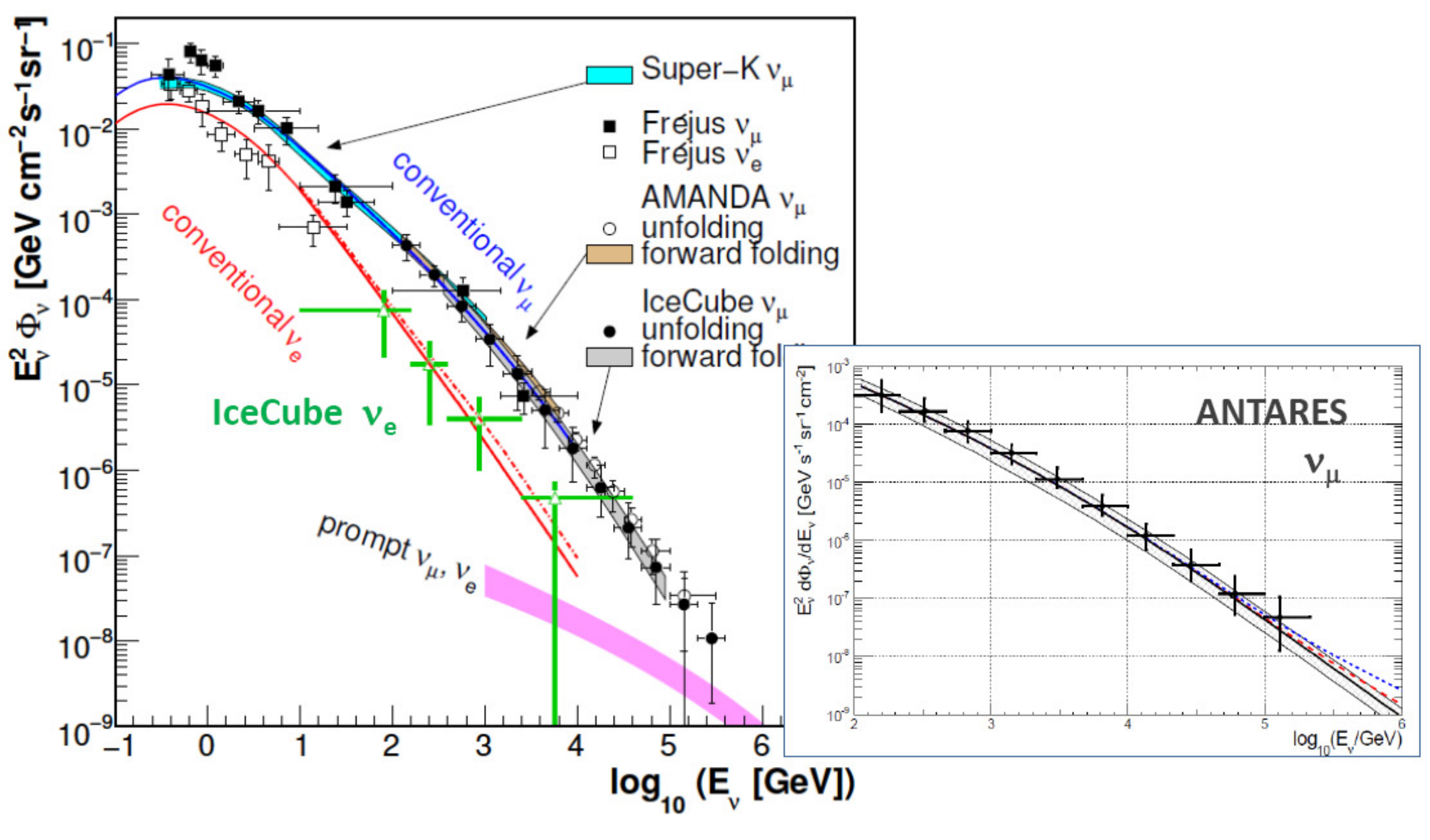}}
\caption
{The energy spectrum of atmospheric neutrinos as measured by various experiments.
IceCube-40 data (left) are given for muon and electron neutrinos and are well
compatible with calculations for conventional atmospheric neutrinos \cite{IC-atm}.
The right side shows the muon neutrino spectrum as measured with ANTARES \cite{Ant-atm}.
}
\label{atmospheric}
\end{figure}

Figure \ref{atmospheric} shows the spectra published by various experiments. 
The data points range up to 200 TeV (ANTARES \cite{Ant-atm}) and 
400 TeV (IceCube-40\footnote{IceCube-40 denotes the IceCube configuration
with 40 installed strings} \cite{IC-atm}) and
are (still!) well compatible with the predictions for conventional atmospheric neutrinos. 
In particular, no significant excess at high energies is observed yet. Improved 
statistics from IceCube (79 and 86 string configurations) will soon allow to test flux models 
for prompt neutrinos or provide evidence for extraterrestrial neutrinos. 
They would show up as a shoulder at some 100\,TeV.

Data points are given for muon neutrinos and electron neutrinos.
The spectrum of muon neutrinos must be deconvoluted from the
measured $dE/dx$ of the recorded muons, taking into account that $a)$ these muons 
will have lost energy before reaching the detector and $b)$ carry only part
of the neutrino energy. For electron neutrinos, the first of these issues does
not play a role: electron showers have lengths of the order of 10\,m, therefore
the events are contained and all energy carried by the electron becomes visible.
On the other hand, electron cascades can hardly be distinguished from
hadronic cascades which form the final state of most $\nu_{\tau}$ interactions and 
of all neutral current interactions. Therefore the $\nu_e$ flux can only be obtained via a
delicate subtraction procedure leading to relatively large errors. Within these
errors, the IceCube data are well compatible with the predictions for conventional
electron neutrinos from the atmosphere.

\begin{figure}[ht]
\sidecaption
\includegraphics[width=6cm]{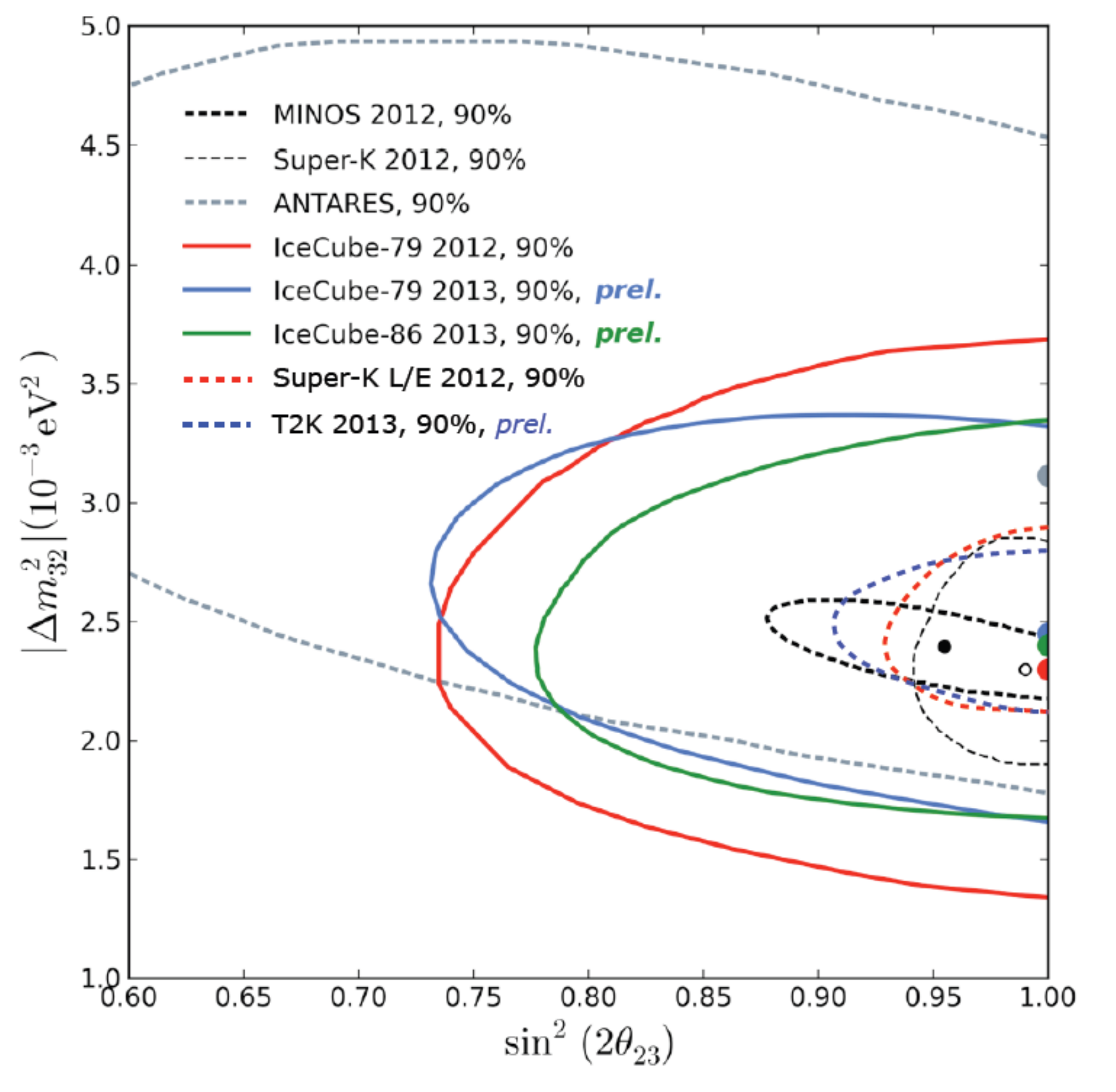}
\caption
{Constraints to the oscillation parameter space from different experiments.
The three independent IceCube analyses are based on one year
of data, two for Icecube-79, one for IceCube-86 (solid lines). The constraints from ANTARES
(gray dashed line) are weaker, those from MINOS, Super-K and
T2K are much stronger. More data and better analysis techniques
will considerably improve the IceCube constraints. 
Figure taken from \cite{deyoung-2013}.
}
\label{oscillations}
\end{figure}

Atmospheric neutrinos also provide a tool to investigate neutrino oscillations. 
The oscillation length scales with $E_\nu$. For distances of the order of
the Earth diameter the first oscillation minimum is at $E_\nu\simeq24$ GeV. 
The suppression of the observed $\nu_{\mu}$ flux is a function of neutrino 
energy and of zenith angle and allows
 to extract the oscillation parameters
$\theta_{23}$ and $\Delta m^2_{23}$. Fig.\,\ref{oscillations} shows
the constraints to the oscillation parameter space from different experiments,
including ANTARES and IceCube/DeepCore \cite{deyoung-2013}.
The constraints from DeepCore data are half the way between those of
ANTARES on the one hand and those of MINOS, Super-K and T2K on the other. Preliminary estimates
show that over a few years, DeepCore can reach a similar sensitivity as
the three latter experiments. The real promise is, however, to further
increase the density of IceCube's core (project PINGU) and to determine
the neutrino mass hierarchy (see section \ref{s-future}).

The great scientific breakthrough of the year 2013 has been obtained
at the high-energy frontier, as will be described in sect.\,\ref{ss-diffuse}. On the other
hand, the progress with neutrino oscillations is similarly impressive.
Going to compete in this field with the best accelerator experiments and with
Super-K was certainly beyond the expectations of most experts.
The perspective to become a main player in fixing the neutrino
mass hierarchy is even more exciting.

\subsection{Steady point sources}      
\label{ss-point}

High-energy cosmic neutrinos may either be identified as accumulation of events
pointing to a particular celestial direction ("point sources'') or as extended
diffuse emission, ranging from a few degrees (as for nearby supernova remnants)
to fully diffuse, essentially isotropic neutrino flux.

Cosmic neutrinos from a given source would cluster around the source
direction. Event statistics has grown and analysis methods have been 
continuously improved over the years,
e.g. by including energy estimators in the analyses, by moving from methods
with fixed widths of the search bin to "unbinned" methods etc. This led
to a tremendous progress.

A nice demonstration of how the sky-maps in  AMANDA and IceCube
have evolved with time can be found in Fig.\,3 of \cite{Karle-2013}.
It starts with meager 17 upward-muon events in 1999 and ends with
some $10^4$ events in 2012: a  fantastic factor-1000 pace in
statistics and in sensitivity (see Fig.\,\ref{point-limits} below)!
In Fig.\,\ref{A+B-sky}
I show another historical sky-map from 2005 which combines data from 
AMANDA (2000-2003) and NT-200 in galactic coordinates.

\begin{figure}[ht]
\sidecaption
\includegraphics[width=6cm,height=3.5cm]{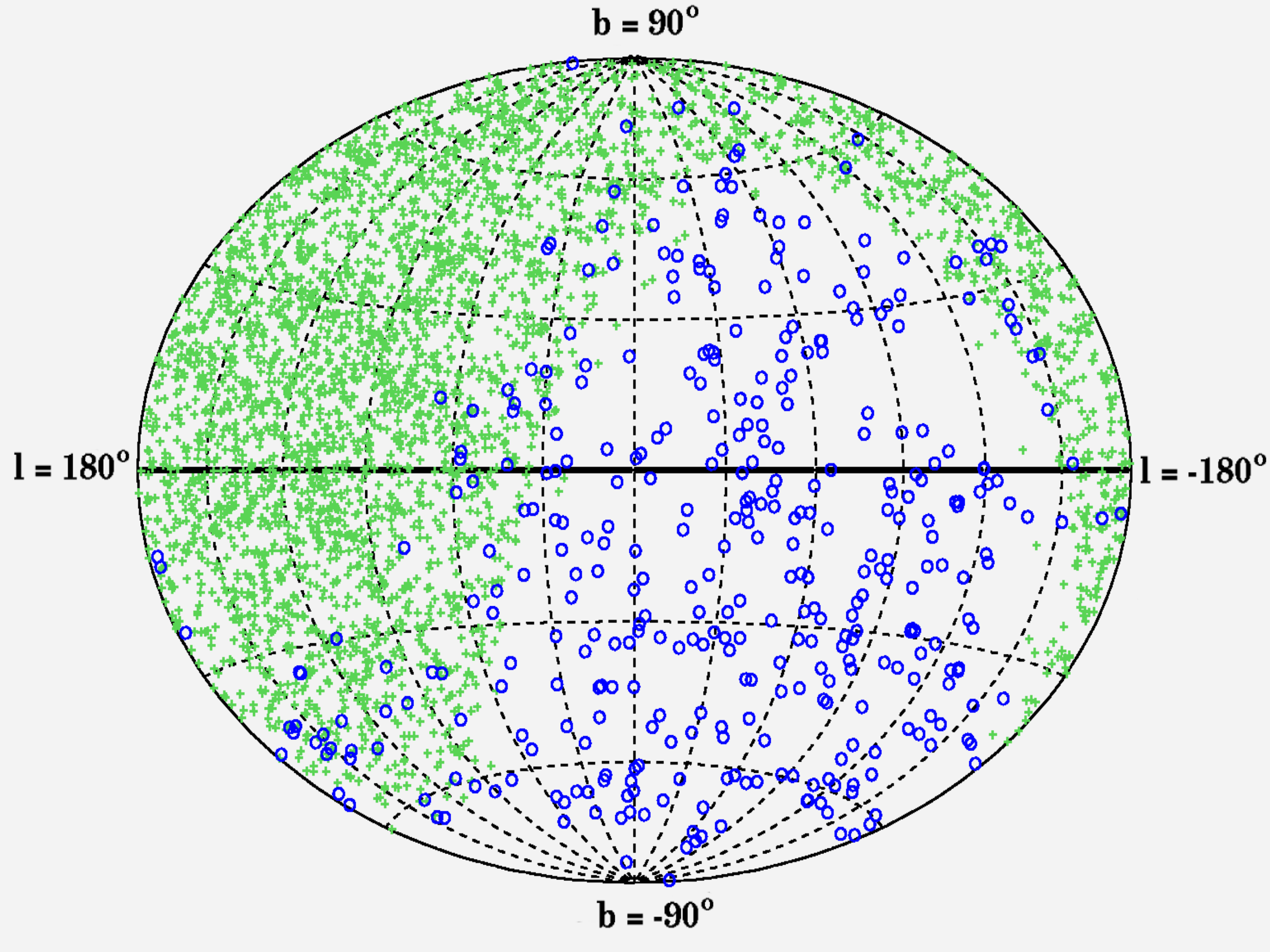}
\caption
{Combined sky-map of upward moving muons recorded
with AMANDA and Baikal NT200 (figure compiled in 2005).
}
\label{A+B-sky}
\end{figure}

\bigskip

\begin{figure}[ht]
\center{
\includegraphics[width=7.5cm]{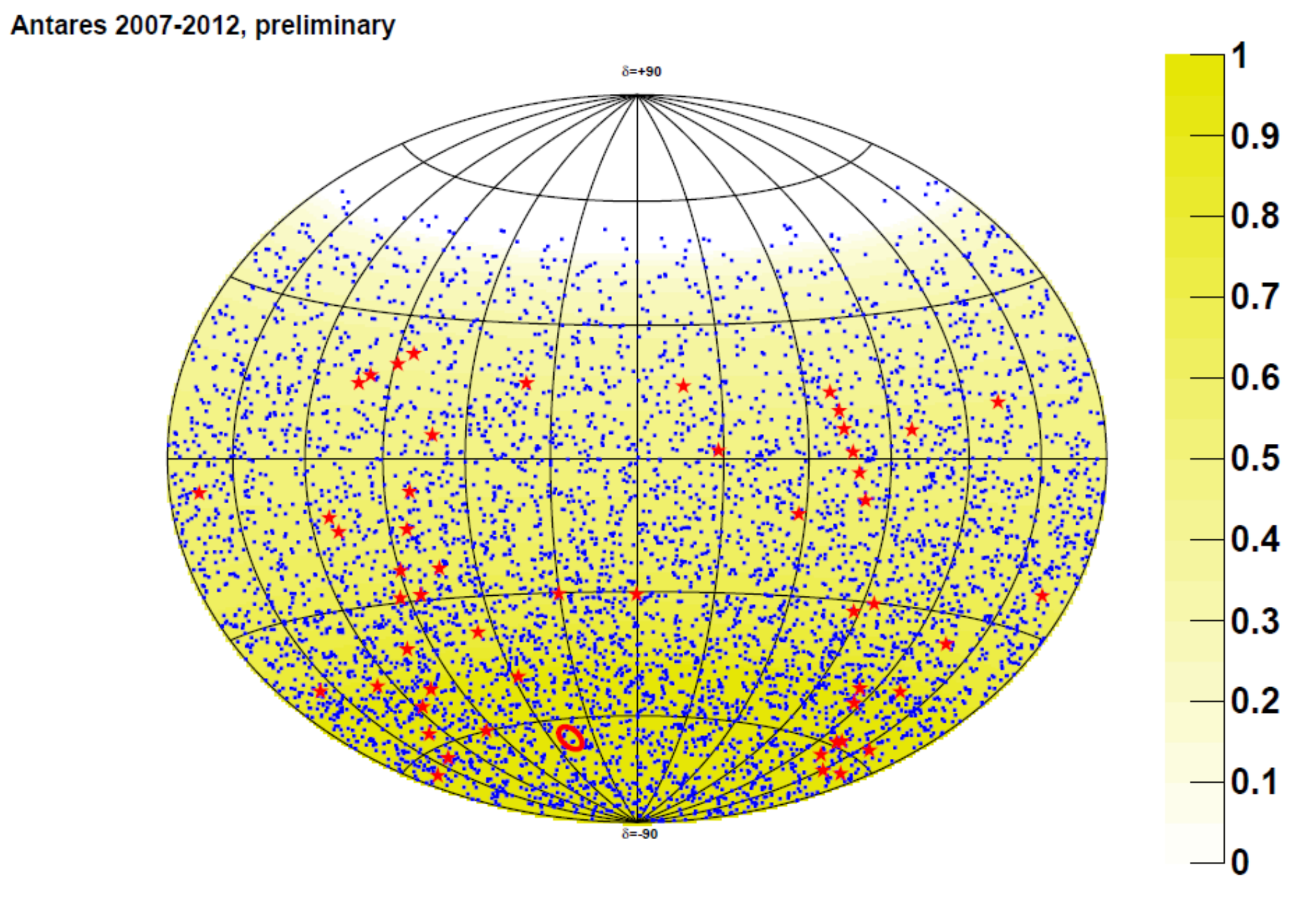}
\hspace{4mm}
\includegraphics[width=6.5cm,height=5.3cm]{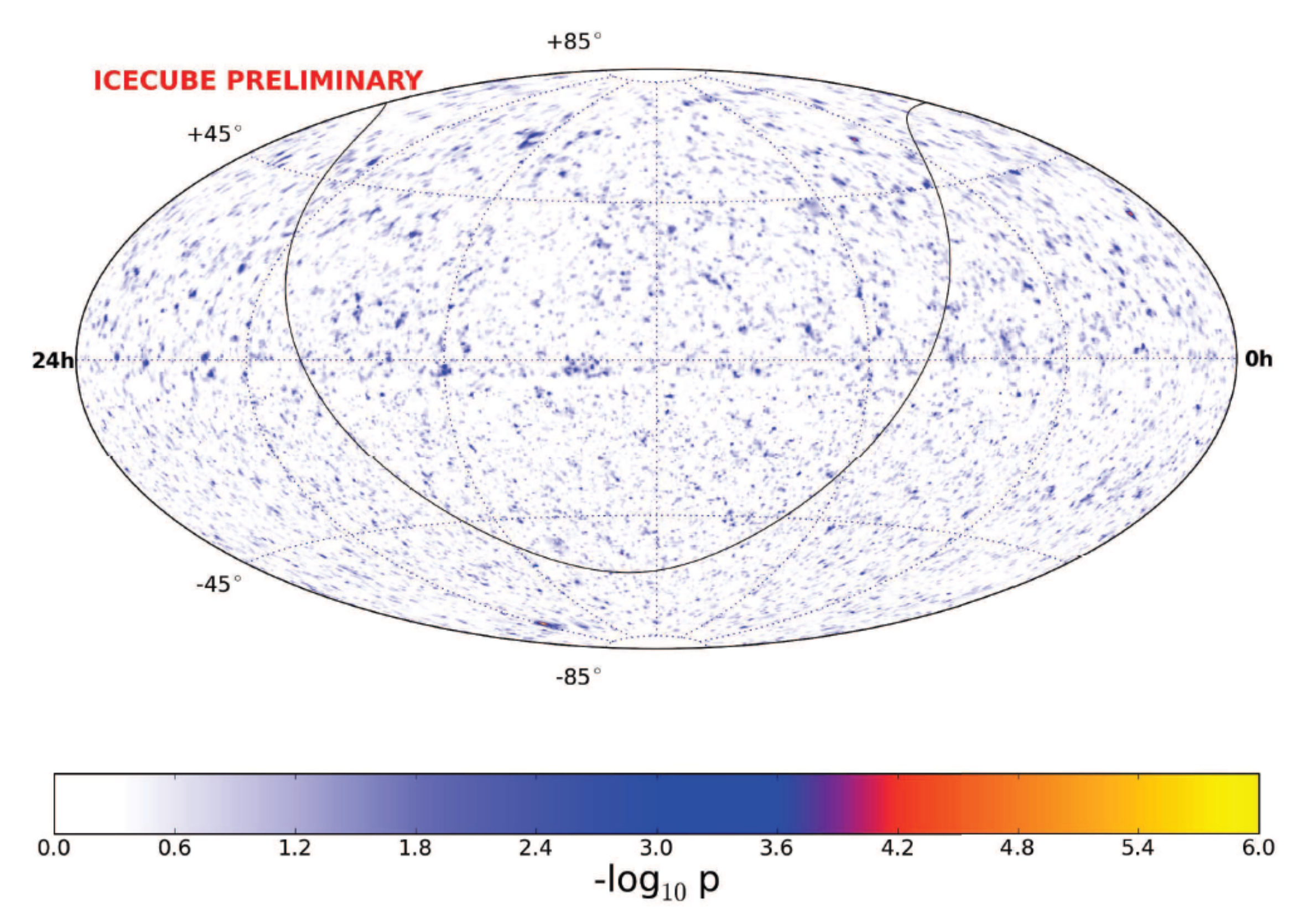}
}
\caption
{Skymaps of ANTARES, 5 years (left \cite{ICRC-ANT}) and IceCube, 4 years
-- 40, 59, 79 and 86 strings, respectively (right, \cite{IC-point}). 
The color left codes the visibility to ANTARES, the color right 
the significance of the excesses.
}
\label{A+I-sky}
\end{figure}

Today, the sensitivity frontier is defined by ANTARES for the Southern sky
and IceCube for the Northern sky (both having their best sensitivity
to point sources by looking down, i.e. through the Earth to the other
hemisphere). Figure \ref{A+I-sky} shows the sky-maps of ANTARES
and IceCube in equatorial coordinates \cite{ICRC-ANT,IC-point}.

ANTARES has used only upward
muons, and the extension of the map to the Northern hemisphere
is due to the fact that ANTARES is not just at the North Pole
and its field of view sweeps over different parts of the sky during one day.
Contrary, IceCube does not change its field of view with the
Earth's rotation. Instead, access to the Southern sky is obtained by 
raising the energy threshold for downward muons -- loosing 
sensitivity at low-energies but keeping it for energies
of PeV or above.  Figure \ref{point-limits} gives the
sensitivities/upper limits obtained from various experiments.
Note that the sensitivity from the first AMANDA analysis 
(AMANDA 10-string array) to that of 4 years IceCube has indeed
improved by more than a factor of 1000!

Naturally, the IceCube sensitivity to a $E^{-2}$ flux from 
the Southern hemisphere is worse than for Northern sources since the
analysis relies exclusively on the tiny high-energy tail of the neutrino flux. 
For unbroken $E^{-2}$ spectra, a cubic-kilometer detector at the
South Pole can compete with a Northern first-generation detector like ANTARES up
to a declination of -$45^\circ$. This means that there is a
broad declination region where the combination of IceCube and ANTARES data will
give a better sensitivity than IceCube or ANTARES alone. Such combined analyses
are presently underway.

\begin{figure}[h]
\center{
\includegraphics[width=10cm]{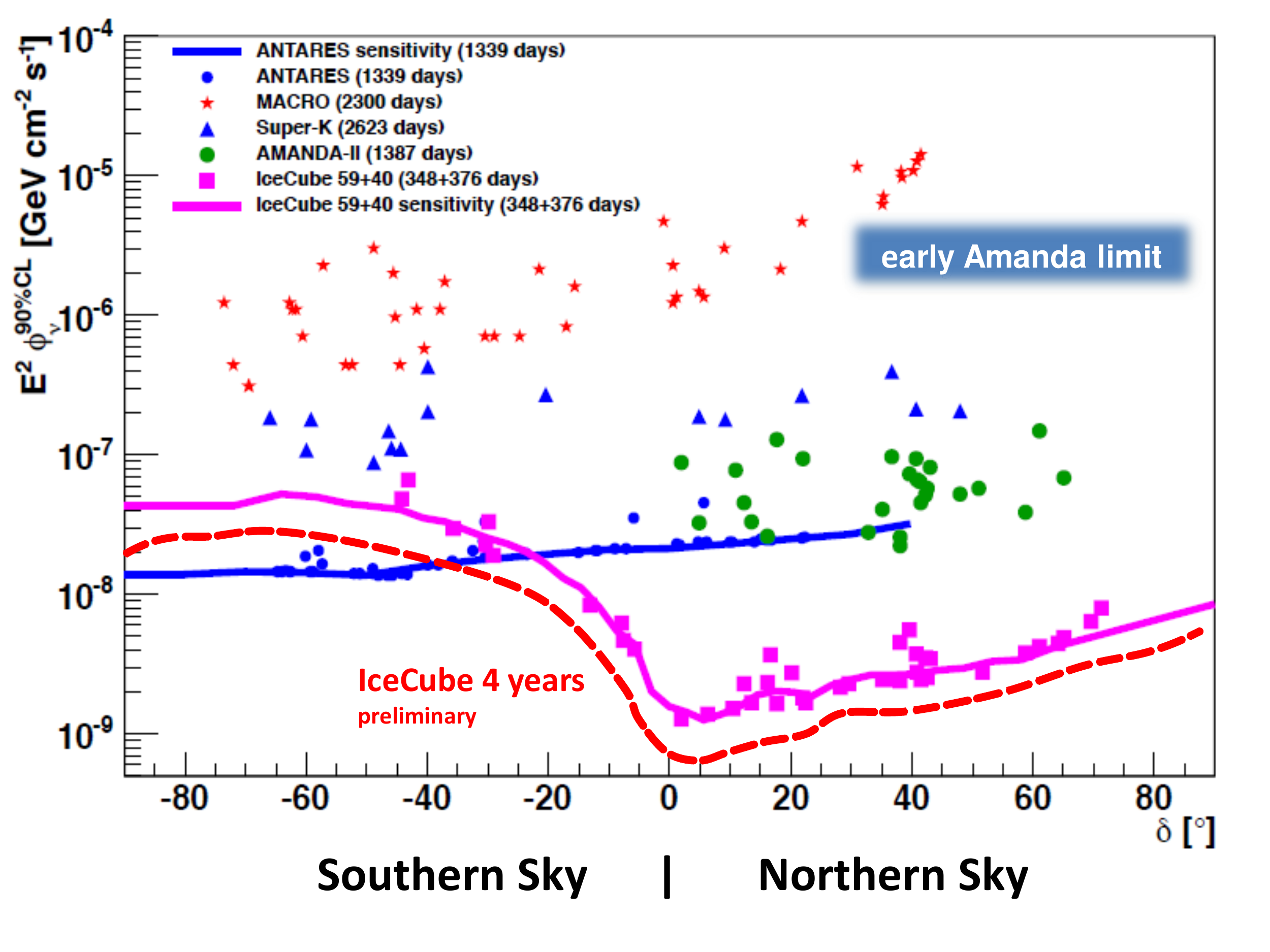}
}
\caption
{Point source neutrino flux sensitivities (median expected limits at 90\%
C.L.) and upper limits for selected sources from various experiments:
Super-Kamiokande, AMANDA, the various stages of IceCube and ANTARES.
See \cite{KS} for references. 
}
\label{point-limits}
\end{figure}

Optimistic flux predictions for some sources are a factor 3-10 below present IceCube limits. 
This does not rule out a discovery with the standard point-source analysis
which focus to TeV energies  -- but we are scraping the discovery
region at best. 

\subsection{Transient sources}           
\label{s-grb}

Many astrophysical sources are known to have a variable flux at different
wavelengths. Examples for such flaring sources are Active Galactic Nuclei, Soft
Gamma Ray Repeaters, and Gamma Ray Bursts (GRB). Binary systems often show a periodic
behavior, as pulsars do. Reduction of the search window to the
time interval of flares or bursts reduces the background from atmospheric neutrinos. 

Here I report results on GRB which 
can last from less than a second to several hundred seconds.
GRBs are now known to be of extragalactic origin and have been 
suggested as dominant sources for the cosmic
rays at the highest energies \cite{WB-1997}. 
In the GRB fireball model, cosmic-ray acceleration
should be accompanied by neutrinos produced
in the decay of charged pions created in interactions between
the high-energy cosmic-ray protons and gamma rays
\cite{WB-1997,Guetta-2004}. 
Both IceCube and ANTARES have searched
for coincidences of neutrinos with GRBs. In
\cite{IceCube-GRB} no coincidences of IceCube events with any of 215 GRBs
were reported, with an expectation of 5.2 coincidences according to \cite{Guetta-2004}.
The 90\% C.L. upper limit was a factor 3.7 below the fluxes predicted by the fireball models.
However, it took not long to show that the predictions turned out to be
too optimistic in several aspects \cite{Winter-GRB}. With revised
calculations, the predicted fluxes are again below the 
published IceCube limits.
Since then, however, data for $\simeq300$ additional GRBs have been analyzed, and
the corresponding sensitivity is at the same level as the revised
predictions. This seriously challenges the hypothesis of GRB
being a dominant source of cosmic rays. 

ANTARES has presented results of a similar search, based on 296 GRBs
in the years 2007-2011 \cite{Antares-GRB}.
No coincidences have been observed. Naturally
the flux limit averaged over all 296 bursts is much worse than that 
of IceCube with its 30 times larger area. 
On the other hand, a discovery of a particularly close GRB, with an
optimum orientation of the jet to Earth and a low Lorenz factor of jetted
matter -- but  outside IceCube's field of view -- cannot be excluded.
This makes continued ANTARES searches important.

\begin{figure}[ht]
\sidecaption
\includegraphics[width=9cm]{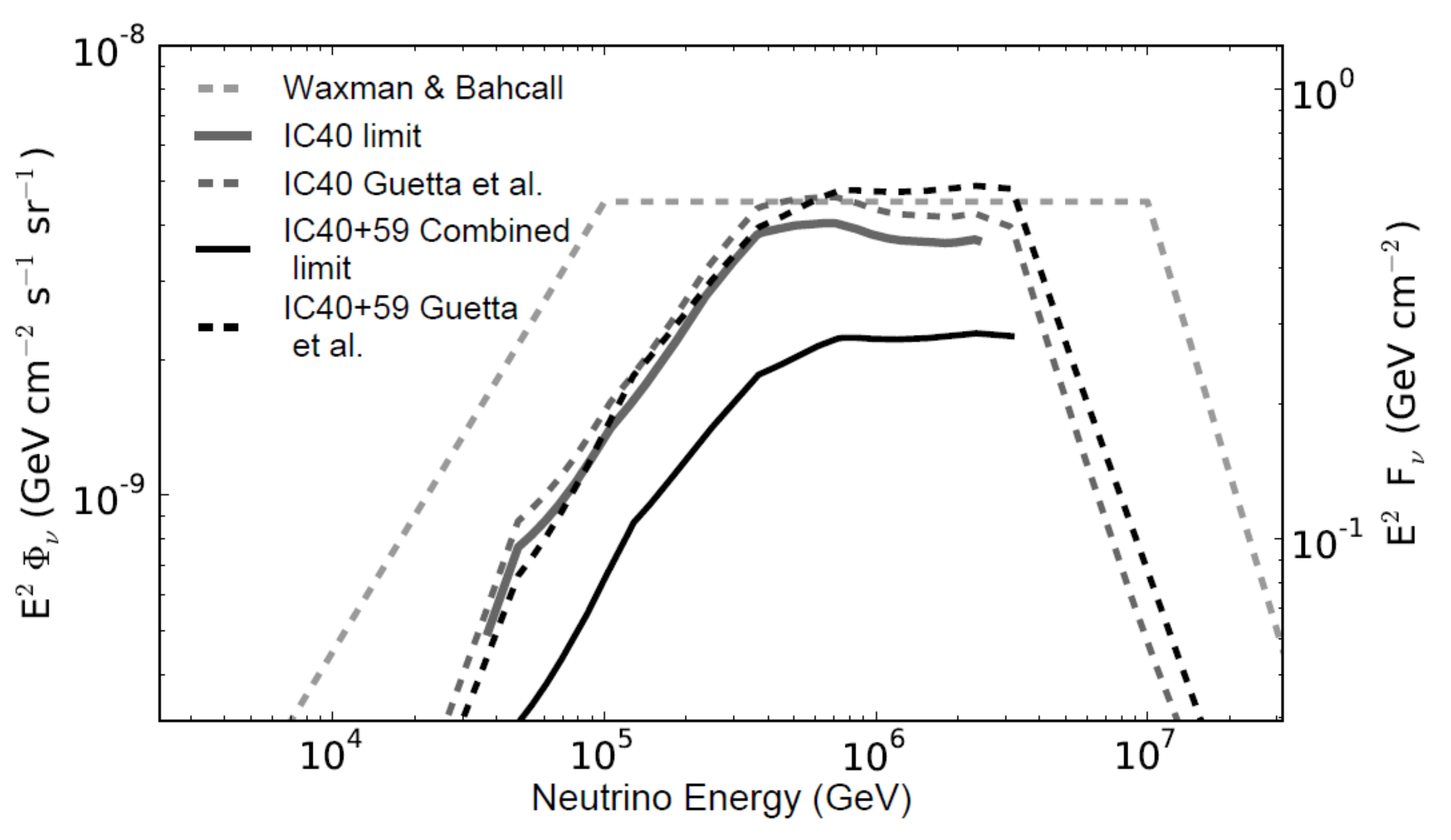}
\caption
{Comparison of initial IceCube results to predictions normalized 
to observed gamma-ray spectra (Guetta et al. \cite{Guetta-2004}) 
and to the high-energy cosmic ray spectrum (Waxmann \& Bahcall
\cite{WB-1997}). 
Figure taken from \cite{IceCube-GRB}. 
}
\label{GRB}
\end{figure}

\subsection{Diffuse fluxes: step to the promised land?}    
\label{ss-diffuse}

It has been shown \cite{Lipari-2008}, that first evidence for
high-energy astrophysical neutrinos is expected from a diffuse
isotropic flux, provided the source candidates 
are dominantly extragalactic and not dominated by a few galactic sources. 
This is a consequence of the fact that neutrinos propagate through
all the universe with negligible absorption, resulting in a 
unresolved flux from all the faint and distant extragalactic sources.

Searches for diffuse fluxes use the measured energy as primary criterion for
separating cosmic and atmospheric neutrinos. A certain distinction of the
one from the other can be obtained from
\begin{enumerate}
\item the ratio between the numbers of cascade events and muons events 
(which is related to the flavor ratio), 
\item the angular distribution,
\item the absence of signals from accompanying punch-through muons
from a possible parent air shower if the neutrinos come from above.
\end{enumerate}

Flavor ratio and angular distribution of cosmic and atmospheric neutrinos
are slightly different. The most important criterion is the energy,
and therefore diffuse analyses critically depend on
a good understanding of the detector response as a function of
energy and on a reliable prediction of the background, most
notably that from prompt atmospheric neutrinos. The latter dominate the high-energy tail of atmospheric neutrinos, and their flux has substantially larger uncertainties than that of conventional atmospheric neutrinos.  

It has been shown in \cite{Kowalski-2005} that the cascade topology has the highest sensitivity for the
detection and characterization of the high-energy excesses. A combination of
all signatures (cascades, downward moving muons, upward moving muons) gives 
the best chance to detect a cosmic diffuse neutrino flux and distinguish if from
prompt atmospheric neutrinos. This is impressively
demonstrated by the recent IceCube analyses.

\begin{figure}[ht]
\center{
\includegraphics[width=5.3cm]{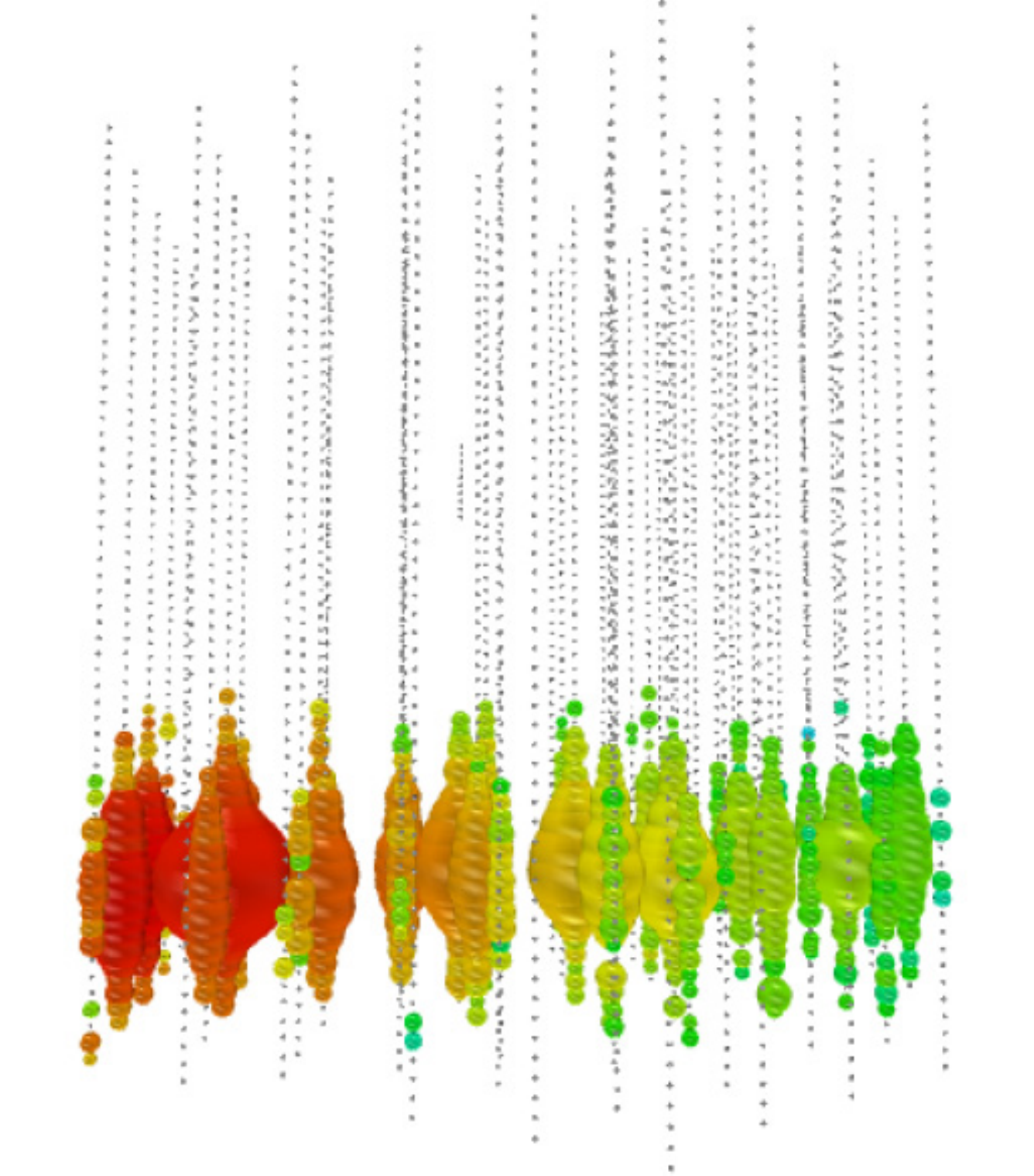}
\hspace{3mm}
\includegraphics[width=8.2cm]{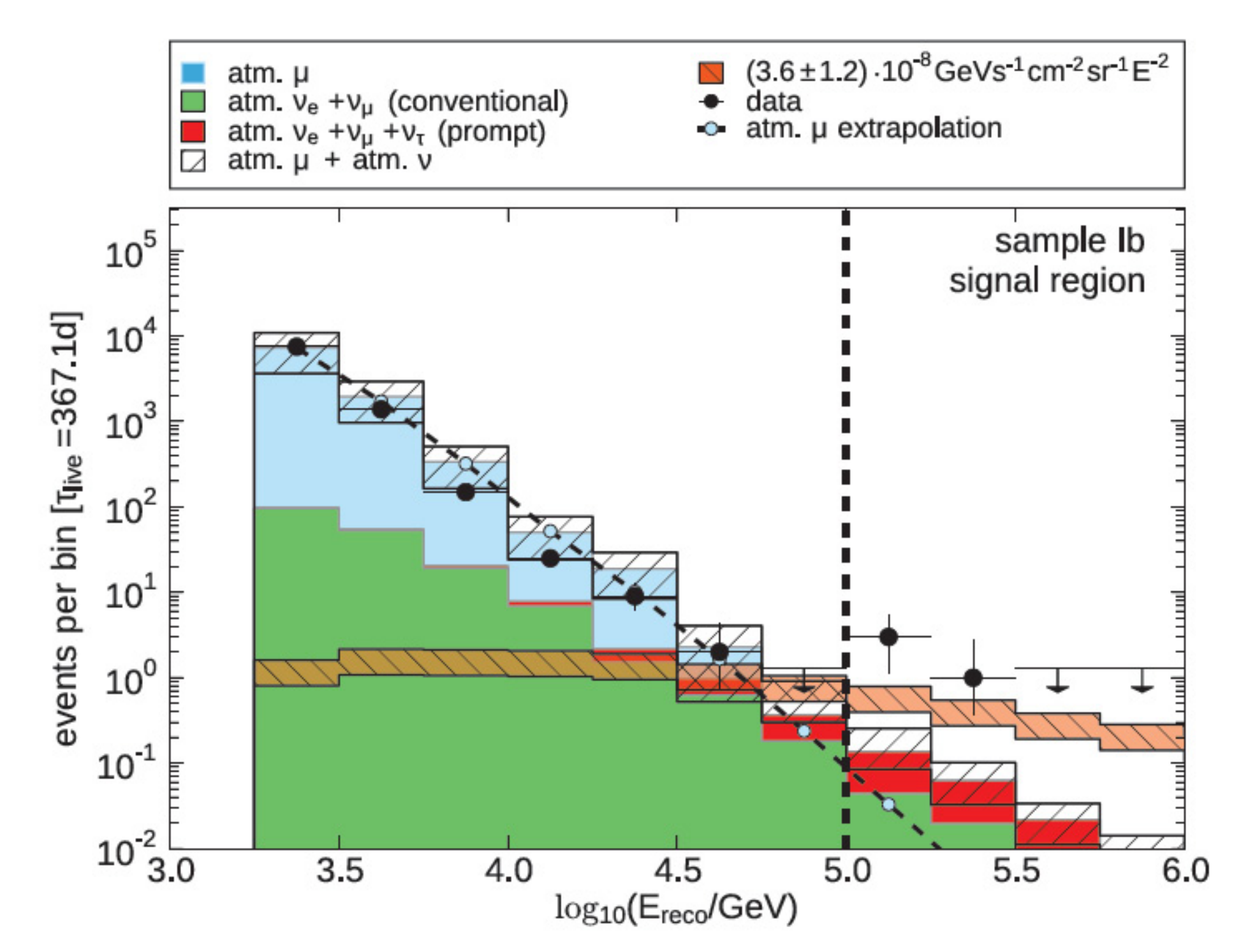}        }
\caption
{Left: 
the highest-energy neutrino event from the IceCube-59 analysis --
a muon entering the detector with about 400 TeV energy, corresponding
to a most likely neutrino energy in the range 500\,TeV-1\,PeV \cite{Anne-2}. 
Right: Reconstructed energy distribution of contained cascades recorded with
IC-40 \cite{Eike}. The white hatched area shows the distribution
of atmospheric neutrinos and muons, including systematic and statistical uncertainties. The orange band indicates the estimate for 
extraterrestrial neutrinos  derived from the HESE analysis described below.
The highest two data points correspond to 4 neutrino events (including one from the
burn-sample which was not used for the significance calculation).
}
\label{Anne+Eike}
\end{figure}

A first candidate for a cosmic neutrino has been obtained in 2012 from 
an analysis of the IceCube-59 data, studying the energy losses 
of about 22\,000 neutrino-generated upward going muons within the detector
\cite{Anne-1}. An event view of the highest-energy
muon, arriving from 1.2 degree below horizon, is shown in Fig.\ref{Anne+Eike},\,left.
The muon enters the detector with an energy of about 400\,TeV, while the most
likely energy of the parent neutrino  is 0.5-1\,PeV. The excess 
w.r.t. to atmospheric conventional and prompt neutrinos is based on this and a second, 
somewhat less energetic muon and has a rather low significance of 1.8$\sigma$.
Translated to an upper flux limit one arrives at
$E^2 \Phi < 1.4 \times 10^{-8}$\,GeV\,cm$^{-2}$\,s$^{-1}$\,sr$^{-1}$ for
an $E^{-2}$ spectrum, or at 3.8 times the value predicted for prompt neutrinos
calculated in \cite{ERS}. Due to the delicate background determination the final
result appeared only recently \cite{Anne-2}.

A 2011 analysis had focused to contained cascades recorded
with the IceCube-40 detector. These cascades deposit all their energy
within the detector volume giving a much clearer correlation to the
neutrino energy than in the case of muons. The mentioned analysis provided 
an excess which was 
compatible with the high background left by filters. 
However, a re-analysis with optimized cuts could
reduce the background considerably while keeping the 4 events 
of the first analysis with
the highest energies (between 140 and 220 TeV) \cite{Eike}.
The energy distribution of the sample is shown in
Fig.\,\ref{Anne+Eike}. The excess has a significance of 2.7\,$\sigma$
over the background of 0.25 events from atmospheric muons and neutrinos.
Again, the delicate and Monte-Carlo-intensive background determination
considerably delayed the release of the final result. A recent similar analysis
of the IceCube-59 data provided a nearly negligible excess and did not
add much significance \cite{Arne}. 

\begin{figure}[ht]
\center{
\includegraphics[width=11cm]{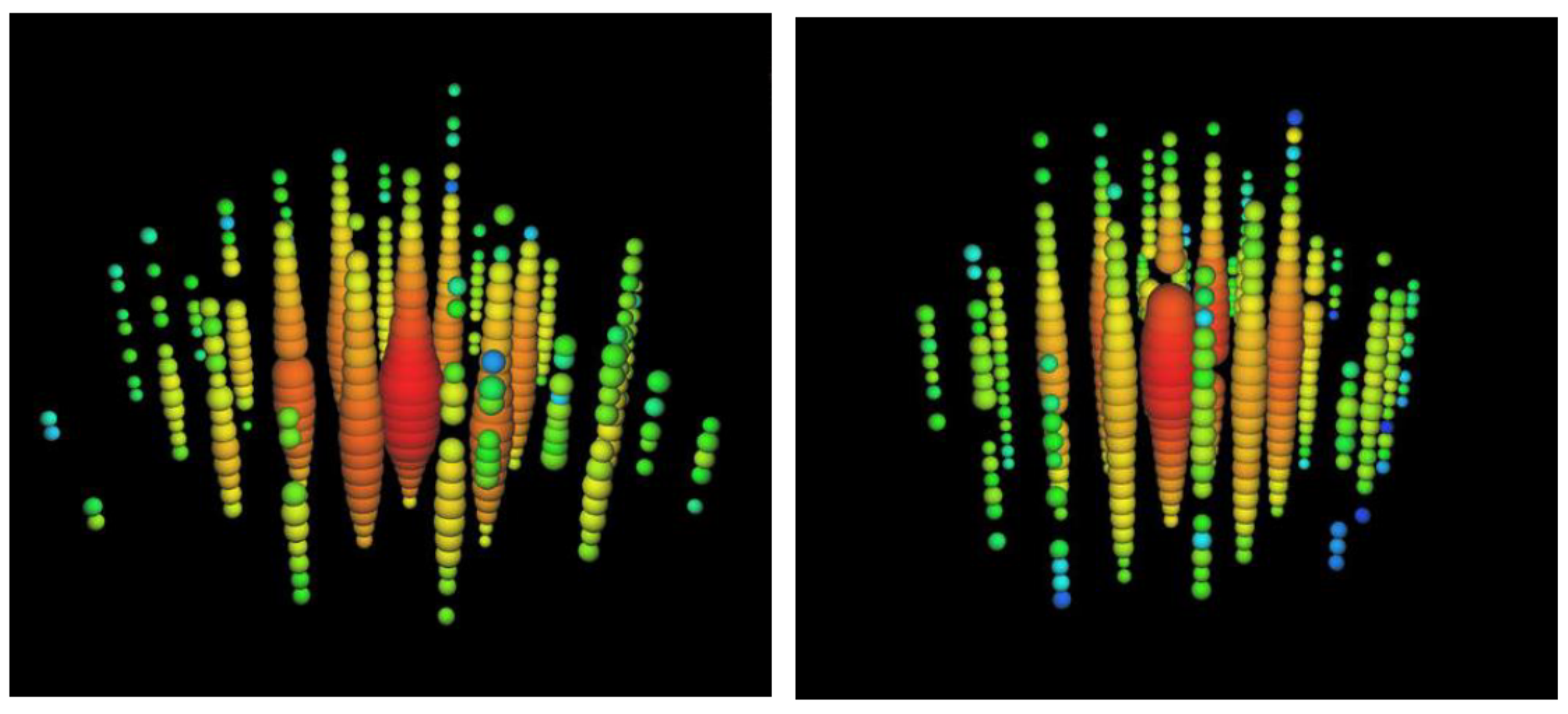} }    
\caption
{The two observed events from January 2012 ("Ernie", left) and
August 2011 ("Bert", right). Each sphere represents a DOM. Colors
represent the arrival times of the photons (red = early, blue = late).
The size of the symbols is a measure
for the recorded number of photo-electrons.
}
\label{E+B}
\end{figure}

A clear step to the PeV scale was made with two events discovered in a search for 
ultra-high energy neutrinos \cite{Ernie+Bert} 
as e.g. expected from GZK interactions of high energy protons with the 
CMB photon field \cite{BZ}. The 
actual energy threshold of the event filter was about 500 TeV.
Data for the search had been taken in 2010 and 2011 with the 79-string and 86-string configurations. 
Two neutrino-induced cascade events passed all filters, with reconstructed energies of 1.14 and 1.04 PeV and were dubbed
“Ernie” and “Bert” (see Fig.\ref{E+B}).
The two events represent a -- still moderate -- 2.8$\sigma$ excess 
over the expectation for atmospheric neutrinos. The sheer energy, however, made them
more promising candidates for cosmic neutrinos than anything found earlier.
Their energy was not high enough for a plausible origin from GZK processes,
the primary motivation for this analysis. They also were considered unlikely to originate
from the Glashow resonance as only about 10\% of such
interactions would deposit 1.2\,PeV or less in the detector in
cascade-like signatures.

\begin{figure}[ht]
\sidecaption
\includegraphics[width=8cm]{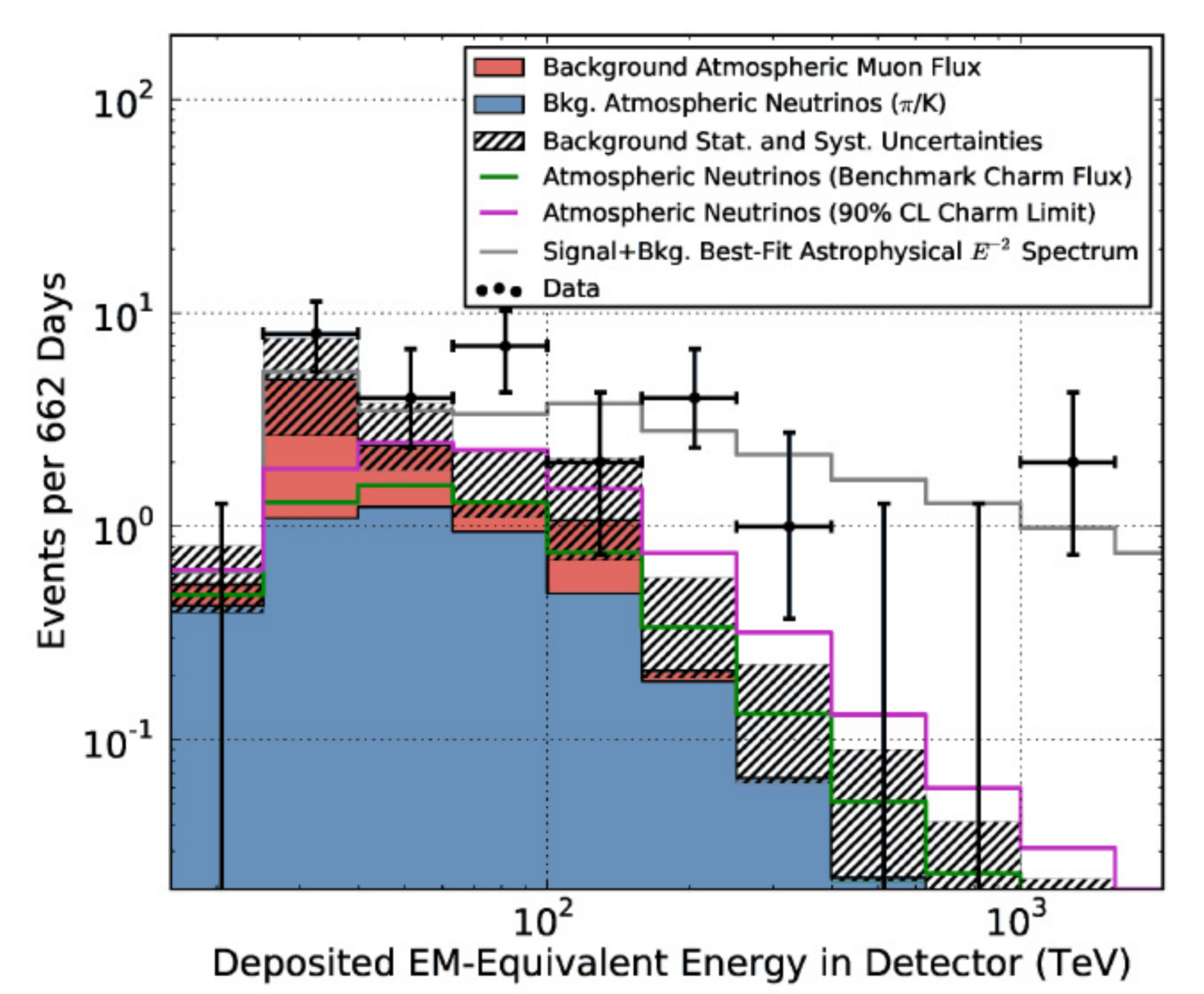}
\caption
{Distribution of the deposited energies of the 28 events compared to model predictions
\cite{Science}.
}
\label{HESE-spektrum}
\end{figure}

Motivated by this result, an alternative analysis of the same data was performed. It constrains the event to start in the inner volume of IceCube (using the outer part as veto layer), and at the same time considerably lowers the threshold compared to the first analysis (down to some tens of TeV). 
New features of this approach included a method for determining the atmospheric muon background
directly from the data and a calculation of the probability that a down-going atmospheric neutrino
will be accompanied by muons which fire DOMs in the veto layer and reject the event as neutrino candidate.
Results have first been  
presented in May 14 2013 at a conference and eventually published in
\cite{Science}. It provides 28 events, with energies deposited in the detector ranging from 
$\simeq30$\,TeV to 1.14\,PeV. 
Figure\,\ref{HESE-spektrum} shows the distribution of the deposited energies. “Ernie” and “Bert” keep their top position in energy. Notably also the events at somewhat lower energies ($\sim30$TeV 
-- 250\,TeV) can hardly be explained alone by atmospheric neutrinos or by muons moving unrecognized from above into the detector. The contribution of such background sources to the total of 28 events is calculated as $10.6^{+5.0}_{-3.6}$ events, resulting in
a statistical significance of $\sim4.1\sigma$. 
The energy spectrum up to 1 PeV is compatible with a an $E^{-2}$ spectrum at a level of
$E^2 \Phi = (1.2 \pm 0.4) \times 10^{-8}$\,GeV\,cm$^{-2}$\,s$^{-1}$\,sr$^{-1}$. The absence of events 
at energies above 1 PeV requires either a cut-off of the $E^{-2}$ spectrum at several PeV, or
a softer spectrum, e.g. $E^{-2.2}$. This flux is slightly below the bound of Waxmann and Bahcall 
\cite{WB-1999} (with the cut-off in disagreement with the assumptions of \cite{WB-1999})
and clearly below that of Mannheim, Protheroe and Rachen \cite{MPR-2001}.

A global fit to all data (standard cascade analysis 40 and 59 strings, upward muon analysis 59 strings
and the starting-event analysis with 79/86 string data) has been applied in \cite{Lars}.
It includes the contribution from prompt neutrinos as a free parameter and a cut cosmic
neutrino spectrum of the form $E^{-2} \cdot \exp(E/E_{\mbox{cut}})$.
The result is compatible with the fit of the 28-event sample alone: 
$E^2 \Phi = (1.0^{+0.8}_{-0.5}) \times 10^{-8}$\,GeV\,cm$^{-2}$\,s$^{-1}$\,sr$^{-1}$ for the
cosmic neutrino contribution,
$E_{\mbox{cut}}$ in the 1-6 PeV range,
and $\Phi_{\mbox{prompt}}$ being $(2.8 \pm 2.0)$ times the prompt flux
calculated in \cite{ERS}.

\begin{figure}[ht]
\center{
\includegraphics[width=8cm]{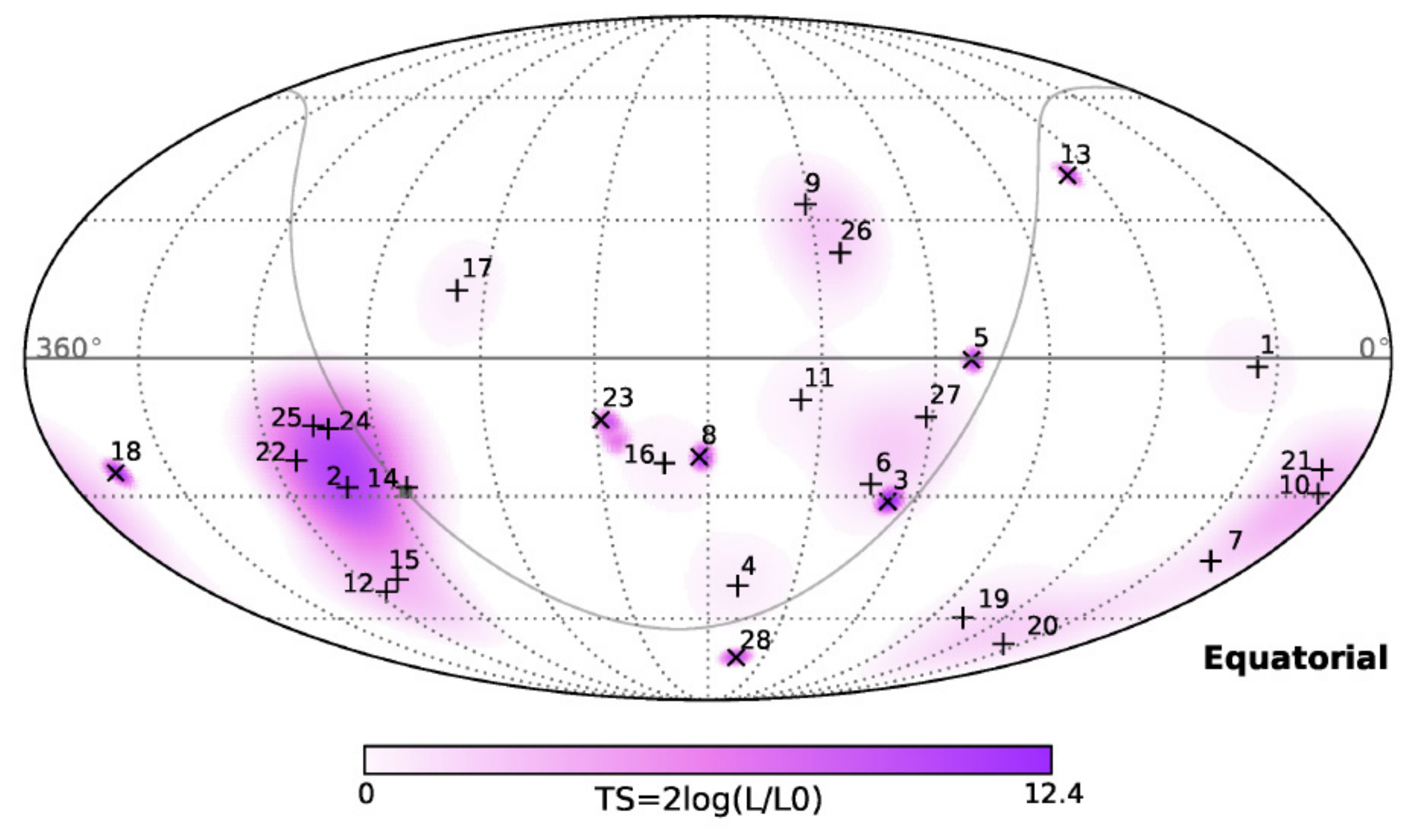}  }
\caption
{Skymap in equatorial coordinates of the 28 events \cite{Science}. The galactic plane is shown as a gray
line, the galactic center as a filled gray square. Best-fit locations for
showers are indicated by crosses and for muons by angled crosses. The coloring
indicates the probability for a point source at that position. 
The cluster close to the galactic center has a chance probability
of about 8\%, i.e. it is not significant.
}
\label{HESE-skyplot}
\end{figure}

\vspace{2mm}
The arrival directions of the 28 events are shown in Fig.\ref{HESE-skyplot}. There is
no significant clustering at any point of the sky, including the intriguing
accumulation close to the galactic center.

Interpreting these results is tempting and has been tried in almost
50 papers which have appeared in 2012 and 2013. A rather complete collection
of references can be found in \cite{Halzen-2013} and
\cite{Anchordoqui-2013}. Explanations include
extragalactic and galactic acceleration processes and decay of superheavy dark matter.
However, the yet limited statistics, the fact that the main significance 
has been obtained with just one special method (the high-energy starting-event analysis) and the
insufficient understanding of the atmospheric (in particular the prompt) neutrino contribution
\cite{Lipari-2013} make it premature to draw clear conclusions. 

Fitting the data from all analyses without any cosmic
contribution is possible if one omits the two PeV events, although with an
extremely high contribution of prompt neutrinos \cite{Lars}. On the other hand, most
down-going prompt atmospheric neutrinos should have been accompanied by
down-going muons -- and these would have been tagged by the veto.

The prompt neutrino signal is related to the spectrum of down-going muons. 
A very high production rate of charm particles
would result in a high-energy shoulder in the muon spectrum. 
A corresponding analysis of IceCube muons is presently underway and will shed
more light on the contribution of prompt neutrinos.

It is also too early to derive source hypotheses which are based on the cascade-to-muon ratio,
firstly since the contamination by background events is different for
cascade and for muon events, and secondly again due to the low statistics.
It is however worth to remind that the high-energy starting-track analysis
suppresses muon-track events in comparison with cascade events in a way which makes the
ratio of 21 cascades to 7 tracks well compatible with a 1:1:1 flavor ratio. 

Soon we will know more. A third event on the PeV scale (christened
"BigBird") has been found when inspecting a 10\% burn-sample of the
second year of IceCube-86 data \cite{Spencer}. A next step is the continuation of the
path started with studying upward muons in IceCube-59 -- the analysis which 
in 2012 had provided
the highest-energy $\sim$400\,TeV muon (see above). A corresponding analysis of the IceCube-79 and
-86 data is underway, and results will likely be released until Summer 2014.
Also the standard cascade analyses of IceCube-40 and-59 will be extended to
data from the later IceCube configurations. 

\section{The future}                      
\label{s-future}

With IceCube, the sensitivity to point sources and to diffuse fluxes
has been improved by more than a factor of thousand when compared to
the situation of the mid nineties. No indications for
extraterrestrial point sources have been found yet, but optimistic source 
models let appear a discovery with IceCube still possible -- with several more years of
IceCube data and improved analysis methods. The first breakthrough, however, has been
obtained when integrating over the full sky, showing
evidence for an extraterrestrial contribution in the diffuse flux.

More than five decades after the first conceptual ideas, and four decades after first 
practical proposals to build high-energy neutrino telescopes, we therefore
may be close to a turning
point. We are likely catching a first glimpse to the promised land of the 
high-energy neutrino universe!

This has important consequences for the future strategy of the field.
For the first time one feels legitimated to give "green light" for building 
on the Northern hemisphere a second detector on the
cubic kilometer scale. 
The danger to build such a second cubic kilometer array and then 
"see nothing" seems obsolete by now.
This does not yet guarantee the identification of point sources, but makes
their discovery more likely than ever before. Certainly, the exact configuration
of large Northern neutrino telescopes should be optimized according to further
findings of IceCube: How important are tracks? How important are cascades? How
good should be the energy and angular reconstruction both cases? etc. But the
way to start building such detectors (in the North as well as an extension of IceCube)
has opened. 

There are two projects on the Northern hemisphere: KM3NeT in the Mediterranean Sea
and GVD (Gigaton Volume Detector) in Lake Baikal.

\begin{figure}[ht]
\center{
\includegraphics[width=7cm]{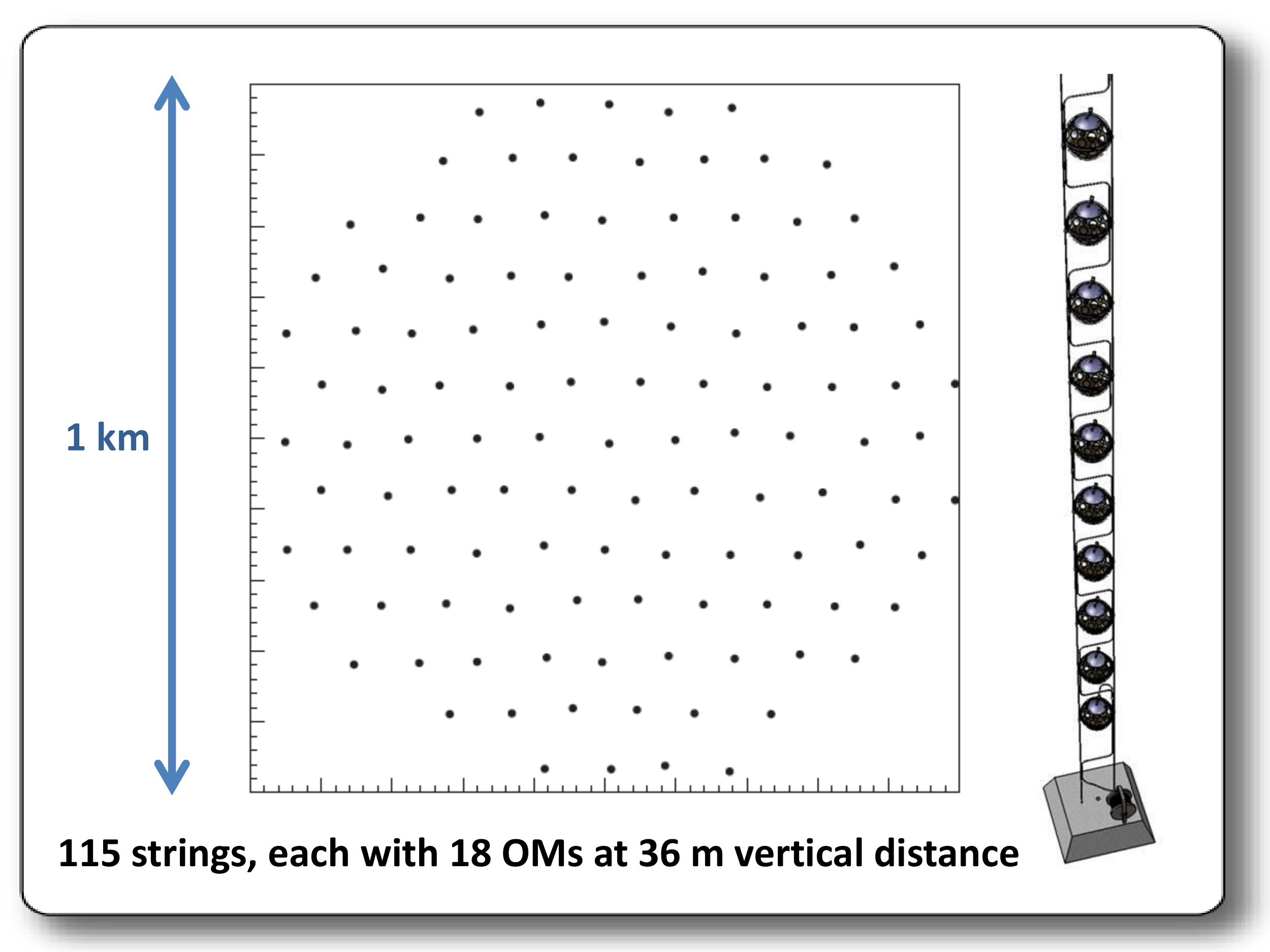} 
\hspace{2mm}
\includegraphics[width=7cm]{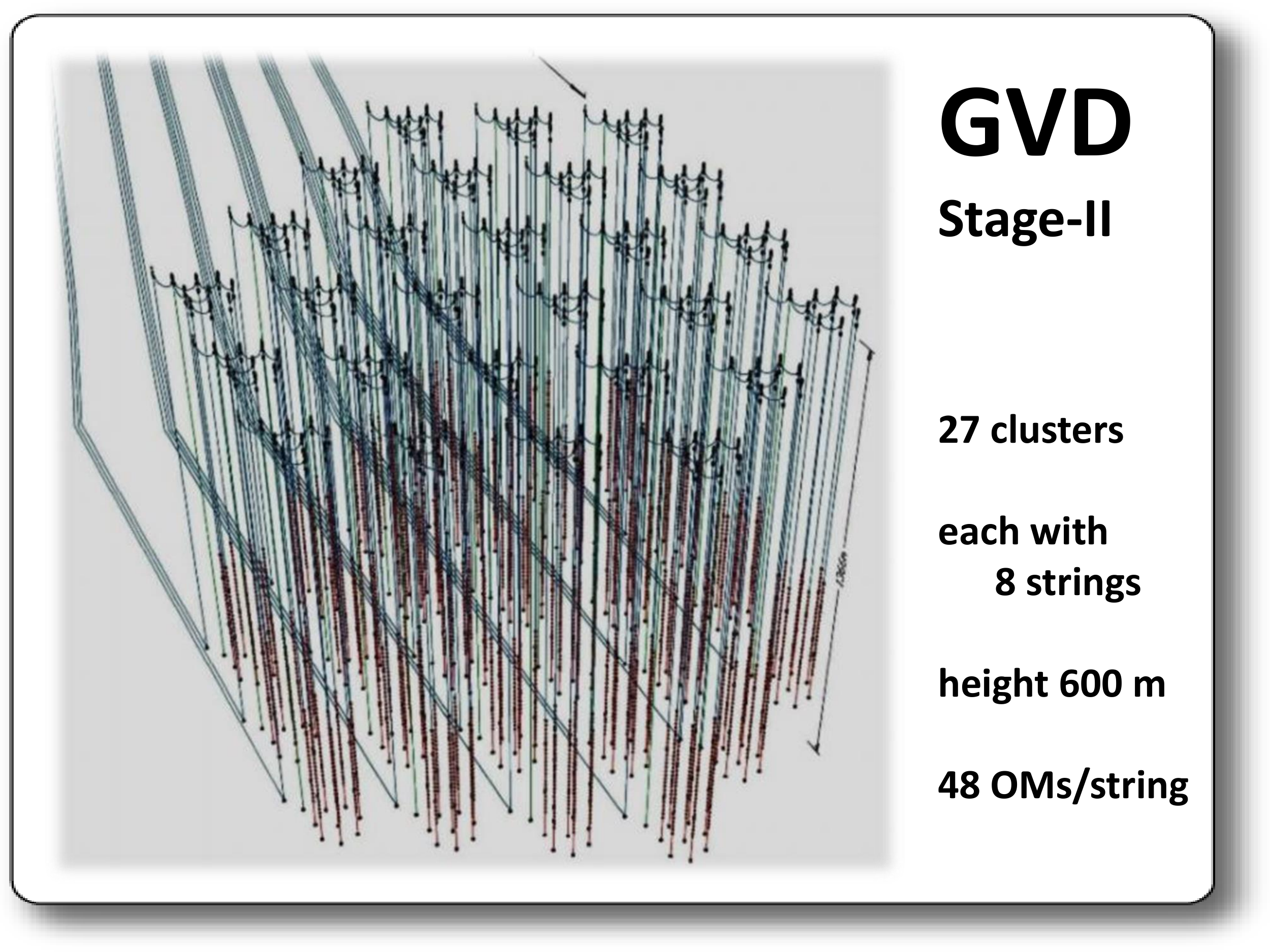} }
\caption
{Left: top view of one of the six envisaged KM3NeT blocks.
Right: artist view of GVD stage-II, with a total volume of about 1.5 km$^3$
}
\label{KM3+GVD}
\end{figure}

KM3NeT \cite{KM3a} will likely be built in the form of several 
separate blocks of the kind shown in Fig.\,\ref{KM3+GVD}\,(left).
After an EU-funded design study 
(resulting in in a Conceptual Design Report (CDR) 
and a Technical Design Report (TDR) \cite{KM3}) and a following
preparatory phase, the KM3NeT
community has recently developed into a formal collaboration. 
They envisage to install
a detector with $\sim$5 km$^3$ volume from 2014 on. The total investment cost
is estimated to be around 225 MEuro. The present plan foresees deployment
of building blocks.
A total of six building blocks could be deployed at three
sites: close to Toulon, close to Sicily, and close to Pylos.

At present, about 40 million Euro have been assigned to prepare 
infrastructures and demonstrator configurations at the French and
Italian sites (KM3NeT Phase-1). The next step will be KM3NeT Phase-1.5
which will 
comprise one or two full cubic kilometer block and allow doing physics at the
level IceCube is doing it now. This would need additional 40-70 MEuro
on top of the assigned 30 million.

In Russia, the Baikal Collaboration plans the stepwise installation of a 
kilometer-scale array in Lake Baikal, the Gigaton Volume Detector, GVD
\cite{GVD}.  It consists of clusters of strings.
Realizing that the originally
planned size of half a cubic kilometer is no longer enough, a four times
larger array is presently being studied, as sketched in
in Fig.\,\ref{KM3+GVD}\,(right).
In the years 2008-2013 the basics elements and an engineering array with a first full-scale string 
and two half-strings have been tested. A Conceptual Design Report is available at
\cite{gvd-cdr}.

\medskip

What about the South Pole?

\medskip
The recent evidence for extraterrestrial neutrinos quite obviously suggests to
extend IceCube to larger volumes and optimize it for higher energies 
(a detector tentatively named DecaCube). First studies
for such arrays have been presented recently \cite{DecaCube}.

\begin{figure}[ht]
\sidecaption
\includegraphics[width=6cm]{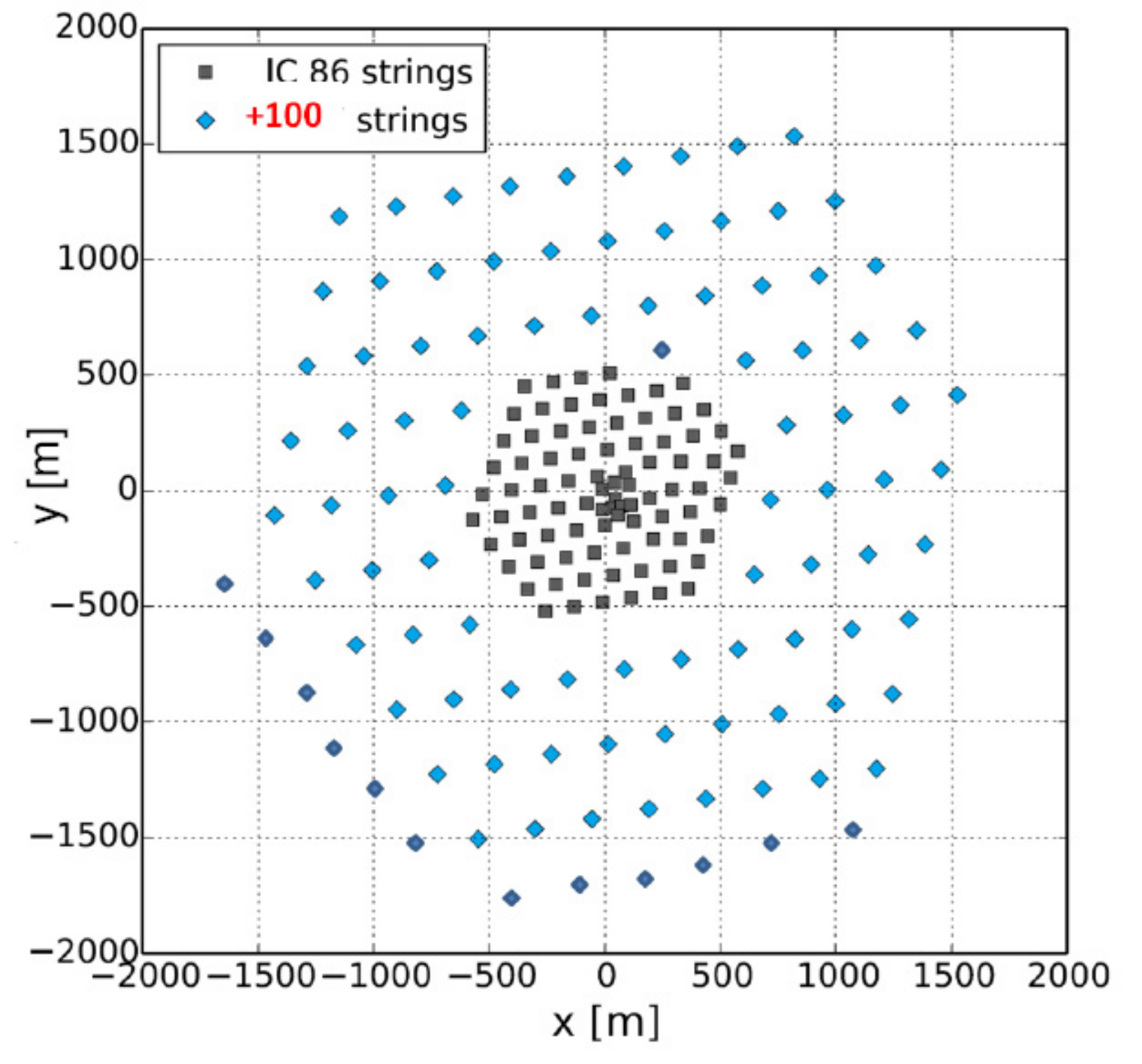}
\caption
{Top view of a possible design of DecaCube,  a high-energy extension of IceCube, 
here with 100 additional strings and a total volume of about 7 km$^3$
\cite{DecaCube}.}
\label{DecaCube}
\end{figure}

One possible configuration is shown in Fig.\,\ref{DecaCube}, with 100 widely spaced strings
arranged around IceCube. With a volume of 7\,km$^3$, this array
would have 3 (7) times the IceCube sensitivity for muon tracks (cascades), 
with an energy threshold of about 10\,TeV \cite{DecaCube}. 
A starting-track analysis like that presented 
in section \ref{ss-diffuse} would yield 4-8 times more signal events than
IceCube-86 (somewhat depending on the achievable background suppression
close to the borders of the array). Since with a next-generation array one must
pretend to clearly identify point sources, optimization to muon tracks
with their superior pointing will be important. 

A further improvement for the selection of down-going cosmic neutrino events could
be achieved by extending the present IceTop array by a factor of $\sim$50, of course
using a much cheaper technology and wider spacing than for IceCube 
\cite{Auffenberg}.
This array could veto atmospheric neutrinos beyond some 10\,TeV with very
high efficiency. 

\begin{figure}[ht]
\center{
\includegraphics[width=9cm]{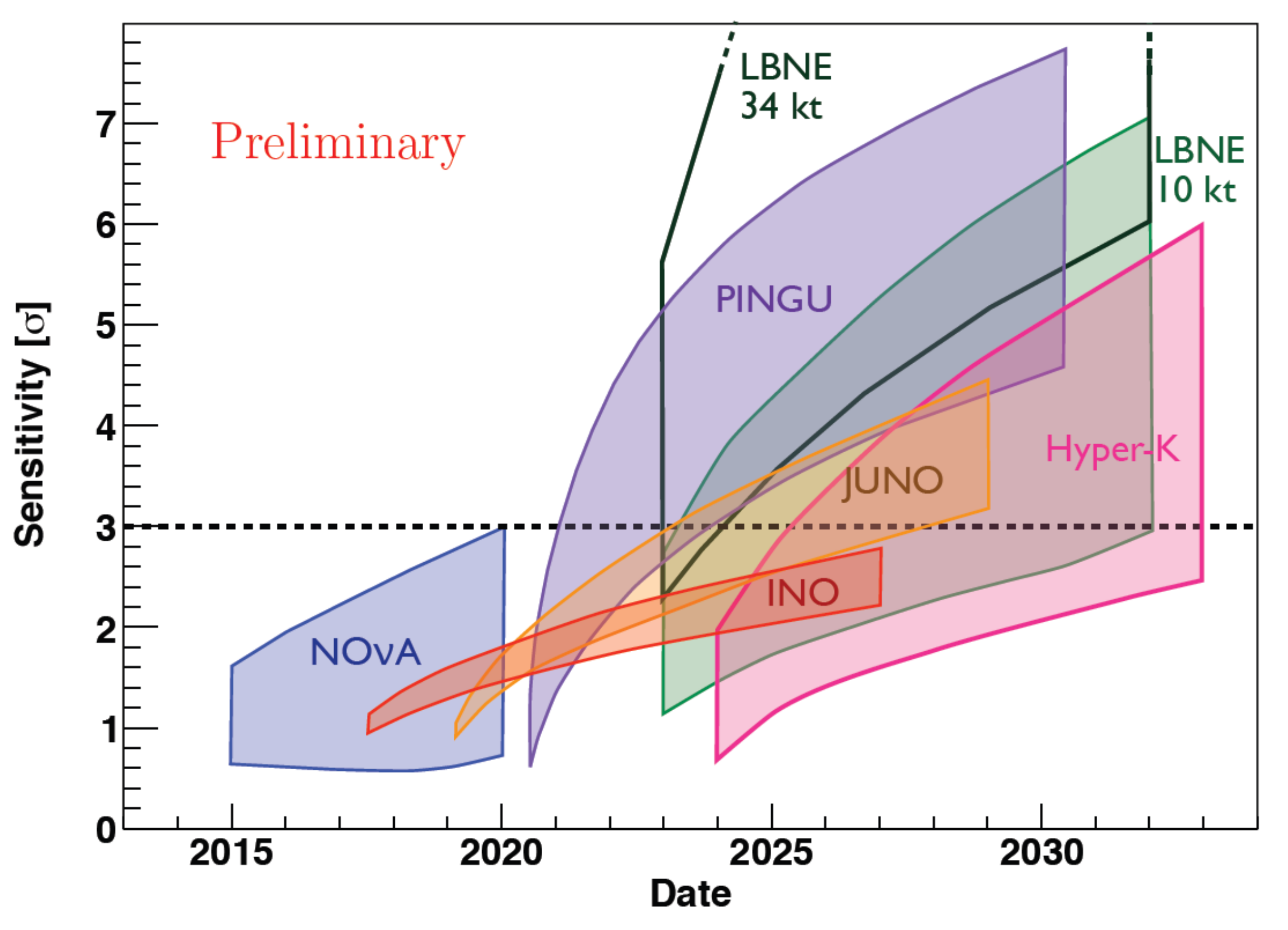}  }
\caption
{Estimated significance for the mass hierarchy to be determined by
several existing or planned experiments, following \cite{Blennow}.
The widths of the bands cover the range of expected sensitivities.
They depend both on the true hierarchy (for Nova and LBNE),
on different true values of the
CP phase $\delta$, on different assumptions on the achievable energy resolution
(for JUNO) and for atmospheric neutrinos on the mixing angle $\theta_{23}$ ranging from the first
to the second octant. The estimated sensitivity for PINGU are those presented
in \cite{PINGU-LoI}, and all other curves are the union of the ranges presented
for the two hierarchies presented in \cite{Blennow}.
}
\label{PINGU}
\end{figure}

The present focus of the IceCube collaboration, however, is on the PINGU project.
PINGU stands for Precision IceCube Next Generation Upgrade
(see the letter of Intent of the IceCube-PINGU collaboration \cite{PINGU-LoI}). 
The primary goal of PINGU
is to determine the mass hierarchy of neutrinos. PINGU exploits the effect of resonance and
parametric oscillations of atmospheric neutrinos propagating through the Earth.
For energies below 10-15\,GeV, these oscillations would cause a pattern in the 
energy-zenith plane which depends on the hierarchy (normal or inverted)
and which could be measurable by PINGU \cite{PINGU}. 
The baseline design of PINGU consists of 40 additional strings, each with 60
DOMs arranged in the inner part of DeepCore. The energy threshold is at few GeV.
Fig.\,\ref{PINGU} shows the estimated significances for the mass
hierarchy as to be determined by several existing or planned experiments.

A similar study is being preformed for the Mediterranean Sea (project ORCA
\cite{ORCA}). It also includes the option to send a pure $\nu$ or
$\bar{\nu}$ beam from Protvino to the ANTARES site \cite{Brunner}.

\bigskip 

The four collaborations -- ANTARES, Baikal, IceCube and KM3NeT -- have recently formed
a "Global Neutrino Network" with the aim to develop a coherent strategy and to exploit
the synergistic effects of cooperation \cite{gnn}.

\section{R\'esum\'e}

The plans for PINGU and ORCA close the circle and
lead this presentation back to its beginning and to the occasion at which it was given -- 
the hundredth birthday of
Bruno Pontecorvo. Pontecorvo would have found the year 2013 as exciting for neutrino
science as our community does. After the determination of the last mixing
angle $\theta_{13}$ in 2011 and 2012, the year 2013 provided a multi-faceted
perspective to determine the mass hierarchy, with projects like
LBNO/LBNE, JUNO and PINGU/ORCA.
 
On the high-energy frontier, we apparently have achieved the long-awaited breakthrough 
and discovered the
first neutrinos from distant cosmic accelerators. The next steps
will be to consolidate the results, to get a better understanding of the
background and of the energy spectrum, and possibly to identify first
sources (individual or stacked). A new window
to the high-energy universe is being opened, and we hope that in the
next decade we can really harvest the fruits of the 40-year work 
in our field!

\end{document}